\documentclass[prd,aps,preprintnumbers,nofootinbib,showkeys,showpacs,amsmath,amssymb,superscriptaddress,floatfix]{revtex4}
\usepackage[dviwindo]{graphicx}
\begin{document}
\title{The Era of Gravitational Astronomy and Gravitational Field of Non-Rotating \\ Single Point Particle in General Relativity}
\author{Plamen~P.~Fiziev\footnote{ E-mail:\,\, fiziev@phys.uni-sofia.bg  and fiziev@theor.jinr.ru} }
\affiliation{
Sofa University Foundation for Theoretical and Computational Physics and Astrophysics,
5 James Bourchier Boulevard,
Sofia~1164, Bulgaria. \\and\\  BLTF, JINR, Dubna, 141980 Moscow Region, Rusia.}
\begin{abstract}
Utilizing various gauges of the radial coordinate, we give a General Relativistic (GR)
description of static spherically symmetric spacetimes with a massive
point source and vacuum outside this
singularity. We show that in GR there exists
a two-parameter family of such solutions to the Einstein equations
which are physically distinguishable and describe the
gravitational field of a single massive point particle with
positive proper mass $M_0$ and positive Keplerian mass $M<M_0$.
In particular, we show that the widespread
Hilbert form of the Schwarzschild solution, which depends only on the
Keplerian mass $M$ and describes Black Holes (BH),
does not solve the Einstein equations with a massive
point particle stress-energy tensor.
Novel normal coordinates for the gravitational field and a new physical
class of gauges are proposed, thus achieving a correct
description of a point mass source in GR. We also introduce a
gravitational mass defect of a point particle and determine the
dependence of the solutions on this mass defect.
The result can be described as a change of the Newton potential
$\varphi_{\!{}_N}=-G_{\!{}_N}M/r$ to a modified one
$\varphi_{\!{}_G}=-G_{\!{}_N}M/
\left(r+G_{\!{}_N} M/c^2\ln{{M_0}\over M}\right)$
and the corresponding modification of the four-interval.
We show that the proper 3D flat space, where these two potentials can be compared,
is the tangent space above the position of the massive point source.
In addition, we present invariant characteristics of the physically and geometrically
different classes of spherically symmetric static spacetimes
created by a point mass. Our findings are important for description of Extremely Compact Objects (ECOs)
studied in relation with possible echoes in Gravitational Waves (GW) recently discovered by the LIGO/VIRGO collaboration.


\end{abstract}
%
%%%%%%%%%%%%%%%%%%%%%%%%%%%%%%%%%%%%%%%%%%%%%%%%%%%%%%%%%%%%%%%%%%%
%\draft
\sloppy
%\scrollmode
%%%%%%%%%%%%%%%%%%%
\newcommand{\lfrac}[2]{{#1}/{#2}}
\newcommand{\sfrac}[2]{{\small \hbox{${\frac {#1} {#2}}$}}}
\newcommand{\ben}{\begin{eqnarray}}
\newcommand{\een}{\end{eqnarray}}
\newcommand{\la}{\label}
\maketitle
%
%%%%%%%%%%%%%%%%%%
\section{Introduction}
LIGO detection of Gravitational Waves (GW) is
the most important finding in gravity
after Sir Isaac Newton's discovery of the gravitational field. Indeed,
Sir Isaac Newton discovered the gravitational field attached to bodies.
LIGO discovered the gravitational field detached from bodies and freely spreading in space.
The last being a qualitatively different novel gravitational phenomenon.

Without any doubts, with the LIGO/VIRGO observations of more than 11 GW events
(see the references at the WEB address \cite{LOGOOPEN}), a new era in fundamental physics  started \cite{Barack2018}.
Finally, the  gravitational astronomy opened a novel window to the Universe and started to give us hitherto unreachable knowledge of the Nature.

These achievements deserve extraordinary careful analysis of all issues, which appear for the first  time  and are accompanied by many uncertainties and unknowns.
Prior ruthless examination of all facts, hypotheses, assumptions, and interpretations, we cannot be sure of what we really see in this newly opened window  to the Universe.

We certainly may refer to the already existing models and theoretical achievements
like General Relativity (GR) well tested in other physical domains
or to a variety of strictly speaking still hypothetical Black Hole (BH) models.

However, the most important in the new situation is the newly appearing opportunity to examine experimentally the theoretical
assumptions, as well as to look for new developments and unexpected physical phenomena.

The used methods for processing the LIGO data were good enough to discover GW without any doubts
but failed to  recover the most important details needed to establish definitely the right theory
of the observed GW and the physical nature of their sources.

For example, these methods were too crude to recover Quasi Normal Modes (QNMs) as fingerprints of BH,
see, for example, \cite{Starobinskij1973,ChandraDetweiler1975,Chandrasekhar1983,Kokkotas1999, Nagar05,Berti06,Berti09,Testing2019,Barack2018}
and a huge amount of references therein.

The used methods also turned out to be not enough to establish the existence
or absence of echoes as fingerprints of Extremely Compact Objects (ECOs)\footnote{
In the last few years the abbreviation "ECO" is used for
Exotic Compact Objects without an event horizon, which are not vacuum BH solutions of gravitational equations.
We use here this abbreviation in a broader sense, including BH in the class of ECOs = {\em Extremely} Compact objects.},
which are not BH \cite{Cardoso16,Mark2017,Cardoso2017,Conklin18,Wang2018,Raposo18,Cardoso2019}.

Between others, GW150914 is the event with the largest signal-to-noise-ratio in the QNMs ringing domain.
None of the methods, recently proposed for finding echoes, have yet found
a significant echo signal in this event \cite{Conklin18} or discarded firmly their existence.

As a result, we still cannot refute firmly many of the alternative theoretical
explanations for the sources of  GW events observed by the LIGO/VIRGO collaboration.

For recent reviews on the above problems, as well as on a series of other ones,
see the talk by Alan Weinstein at the First LIGO Open Data Workshop 2018 \cite{LOGOOPEN} and a large review paper on the general situation in gravity \cite{Barack2018}.

The most recent observation of the shadow of the ECO
at the center of the elliptic galaxy M87 proves the extremal compact character of the central object \cite{EHT}
but it does not give any indication of the presence or absence of an event horizon of this object.

In the present paper, we consider in a mathematically  correct way the most compact object - a non-rotating massive point particle in GR.

We dub this subject "the Schwarzschild problem", since the first attempt to solve it was made by Schwarzschild in the famous paper
\cite{Schwarzschild} titled "On the Gravitational Field of a Mass Point
according to Einstein's Theory", as early as in 1916.

It is physically clear that this model is a good approximation for description of
any spherically symmetric non-rotating matter object of finite dimension,
as seen from a large-enough distance.

This means that if one measures physical quantities like masses, mass defects, etc,
of bodies of finite dimension, being at large-enough distance from the body,
the measured values are the same as for the corresponding point particles.

The correct mathematical description of "to be simplest" the Schwarzschild problem in GR is
not a trivial issue. It turns to be essential for solution of some of the above mentioned problems
related with ECOs, binary and multiple systems of ECOs, etc, in the era of GW astronomy.

Our approach to the problem is based on our previous papers \cite{Fiziev2003,Fiziev2004a},
which have already a more than sixteen years history.
Those papers were reported at various seminars and discussed with many people privately.

Before the discovery of the GW, these papers did not attract adequate attention of the general audience,
which is happy with the most-well-studied Hilbert form of the Schwarzschild solution (see details in the text below.).

However, in the last few months one can see increasing interest in the paper  \cite{Fiziev2003}
in the scientific nets like Academia, Research Gate, etc, see also \cite{Petrov2018}.
This stimulated the actualization of this paper and its adaptation to the present-day hot topics of fundamental physics.

The most significant finding of this approach is a novel two-parameter-family
of exact solutions of the Einstein equations for a massive point particle,
which are regular in the standard domain of the radial variable $r \geq  0$.

As all other matter objects, in GR these supercompact ones have no event horizon and are defined by
positive proper mass $M_0>0$\footnote{The proper mass $M_0>0$ is defined, as usual, as the mass of the physical system with gravitational interaction turned off. See, for example, the Landau-Lifshitz book in \cite{books}. In the neutron star physics the term "barion" mass is used often for the proper mass. This is not precisely the case since the proper mass can be associated not only with barions.} and positive Keplerian mass $M<M_0$ of the matter-point-particle with mass-defect $M_0-M>0$,
or gravitational binding energy $E_{binding}=\left(M_0-M\right) c^2>0$\footnote{Further
one we use units in which the velocity of light is $c=1$.}.

The mass-defect of the matter bodies is described more conveniently by the mass ratio $\varrho=M/M_0  \in (0,1))$.
It was introduced in the Schwarzschild problem for the first time in \cite{Fiziev2003}.

The ratio $\varrho$ is of fundamental significance for all ECOs that are not BHs, and must be included in their physical description.
In particular, it restricts from below the physically admissible values of the luminosity distance
$\rho \in [\rho_{min}, \infty)$ of the ECO.

The minimal luminosity distance $\rho_{min}$ is a coordinate independent quantity that defines
the luminosity of the point object at a small distance from its center.
In GR, the center of the such ECO is placed in position with luminosity radius value $\rho_{min} > \rho_G$.
Hence, the point sources with the same intensity and larger $\rho_{min}$ are darker.
In GR, when one approaches the position of a massive point source, its luminosity goes to a finite limit $\sim 1/\rho_{min}^2$.

Let $\rho_G=2 G_N M$ be the gravitational radius of the massive point particle with the fixed Keplerian mass $M$. Sometimes this radius is  called also the Schwarzschild radius. Here and further on $G_N$ is the Newton gravitational constant.

The gravitational radius appears for the first time in the model of invisible superheavy stars, invented by John Michell (1783) and Pierre-Simon Laplace (1793)
\cite{ML} in the Newton gravity. Here we present their results in modern notation:

Let light photon starts move with velocity $c=1$ in the radial direction from the surface of a spherical star with the radius $R_*$,  and let $R_{max}$ be the radial distance to the point, where the photon stops and turns back to the star, due to gravitational pull.

Then, in the Newton gravity, the relation
$$1/R_*=1/\rho_G+1/R_{max}$$
takes place. It is a direct consequence of conservation of the photon energy in the Newton gravitational field.
The mean density of the star mass is
$$\mu_*={\frac 1 \kappa}{\frac {3(n+1)^3}{\rho_G^2}},$$
where $\kappa={\frac {8\pi G_N}{c^2}}$ is the Einstein constant and $n=\rho_G/R_{max}$.

For n=1, the radius of the superheavy star is $R_*=\rho_G/2$ and the light can spread only inside the gravitational radius $\rho_G$.

Nevertheless, the Mitchell-Laplace superheavy star, being made of some matter, is not a BH, which is a vacuum solution of the Einstein equations \eqref{Einst}.

For example, if the central ECO in our Galaxy is a Mitchell-Laplace superheavy star, its mean density will be $\sim 9 \times 10^{3}{\frac{g} {cm^3}}$, i.e., about 0.01 of the density of a white dwarf. In the same case, the central ECO in M87 will have a lower mean density $\sim 4 \times 10^{-3}{\frac{g} {cm^3}}$, i.e., about 4 times the density of the water, or about the main density of a typical planet as the Earth.

We do not think that the Mitchell-Laplace superheavy star model gives a correct description of the observed real superheavy objects. Of course, the correct model has to be based on GR or on its proper relativistic generalization.

Here we only wish to stress three important facts: 1) As a limiting case of GR, the Newton model contains some part of the physical truth, as well as the corresponding quantities, which appeared in GR. 2) The Mitchell-Laplace superheavy star model is not a predecessor of BH, which are a vacuum phenomenon in GR. The Mitchell-Laplace model is just a model of  {\em matter} body, which is invisible in the electromagnetic spectrum, and does not describe  a hole in the spacetime. 3) The Mitchell-Laplace superheavy star model does not require some more exotic matter, than the one, we find in GR models of the observed objects.

In the present paper, we shall show  that in GR description of a massive point particle with fixed Keplerian mass $M$ and minimal value of the luminosity radius $\rho_{min}$,  its proper mass $M_0$ can be represented in the form
$$M_0=M/\sqrt{1-\rho_G/\rho_{min}}\quad \Leftrightarrow \quad \rho_{min}= {\rho_G / \left(1-\varrho^2\right)}.$$

These basic formulas relate the minimal luminosity distance $\rho_{min}>\rho_G$ with the mass ratio $\varrho\in (0,1)$ of the point mass. According to those relations, the massive point sources of the same intensity but with a bigger $\varrho$ (i.e. of a smaller mass defect $M_0-M>0$)  are darker.

Above the corresponding minimal value $\rho_{min}$ of the luminosity distance,
the Birkhoff theorem is strictly valid for all solutions of the Schwarzschild problem.

For BHs, being vacuum solutions of Eq.\eqref{Einst} of the finite Keplerian mass $M>0$,  proper mass $M_0\equiv 0$ has never been introduced.
The Schwarzschild BH solutions, as vacuum solutions, have no defect of mass and in this case the ratio of masses $\varrho=\infty$ is a senseless quantity.

As we shall see in detail, in GR the BH solutions form a specific gauge sector of spherically symmetric solutions. This gauge sector is completely different from the one of massive point sources of gravity in GR, both physically and mathematically.

In the last few years, a special kind of ECOs is intensively discussed in the literature, see the review paper \cite{Cardoso2019} and a large number of references therein.

These ECOs were introduced as alternatives to BH and, especially, to study possible echoes in GW during the QNM-ringing phase of binary merger.  For this purpose, the ECOs need to have a luminosity radius well-below the luminosity radius of the photon sphere, i.e.,  $\rho_{min}<\rho_{ps}={\frac 3 2}\rho_G$.

For appearance of observable echoes, the condition $\rho_{min}=\rho_g (1+\epsilon)$ with $0<\epsilon \ll 1$
is preferable \cite{Cardoso16,Mark2017,Cardoso2017,Conklin18,Wang2018,Raposo18,Cardoso2019}.
Then, $\varrho\sim \sqrt{\epsilon}$ and we obtain $M_0\sim M/\sqrt{\epsilon} \gg M$.
Hence, the binding energy $M_0-M\sim M (1/\sqrt{\epsilon}-1) \gg M$. Such ECOs are strongly bounded by gravity and  are very stable.

When the Keplerian mass $M>0$ is fixed,  for $\epsilon \to 0$ we obtain a limiting case with $\rho_{min}=\rho_G$,  $\varrho =0$,  infinite $M_0$ and infinite binding energy. This is not a BH, since at $\rho=\rho_G$ we have a reflecting mirror described by the corresponding jump ($\sim M_0\to \infty$) of the metric  for $\rho<\rho_{min}$.

The value $\varrho=1$ indicates the absence of mass defect and is not admissible in GR.
In this case, we have a non-physical absolutely dark noncompact object with $\rho_{min}=\infty$ and zero binding energy.
This object is gravitationally not stable.

We present here an updated and corrected version of the paper \cite{Fiziev2003}, pointing out
some of its relations with modern hot topics of fundamental physics.

The basic character of the Schwarzschild problem for gravitational physics
prompts us to keep the careful and detailed character of our investigation.

The long history of the Schwarzschild problem requires paying attention to different mathematical and physical
achievements of a large number of authors, ignoring possible misunderstandings and mistakes
in many of the cited papers that become obvious when they are assessed from a modern point of view.

\section{Gauges in General Relativity}

The Einstein equations:
\ben G^\mu_\nu = \kappa T^\mu_\nu,\quad \mu,\nu=0,1,2,3;
\la{Einst}\
\een
determine the solution of a given
physical problem up to four arbitrary functions, i.e., up to a
choice of coordinates. This reflects the well-known fact that GR
is a specific kind of gauge theory.

According to the standard textbooks \cite{books},
the fixing of the gauge in GR in a {\em holonomic}
frame is represented by a proper choice of the quantities
\ben
\bar\Gamma_\mu\!=\!-{{1}\over{\sqrt{|{}^{(4)}g|}}}g_{\mu\nu}
\partial_\lambda\left(\sqrt{|{}^{(4)}g||}g^{\lambda\nu}\right),
\la{Gammas}
\een which
emerge
when one expresses the 4D d'Alembert operator in the form
$g^{\mu\nu}\nabla_\mu\nabla_\nu=g^{\mu\nu}
\left(\partial_\mu\partial_\nu-\bar\Gamma_\mu\partial_\nu\right)$.

We shall call the change of the gauge fixing expressions (\ref{Gammas}),
{\em without} any preliminary conditions on the analytical behavior
of the used functions,  gauge transformations in a {\em broad} sense.
This way we essentially expand the class of
admissible gauge transformations in GR.

Of course, this will alter
some of the well-known {\em mathematical} results in the commonly
used mathematical scheme of GR.

We think that a careful analysis of such a wider framework for the gauge
transformations in GR can help us to clarify some long-standing {\em physical}
problems.

After all, in the physical applications,
the mathematical constructions are to reflect in an
adequate way the properties of the real physical objects.

It is well known that in the gauge theories we may have different
solutions with the same symmetry in the base space.

In general, these solutions
belong to different gauge sectors owning to different geometrical,
topological and physical properties.

For a given sector of solutions, there
exists a class of {\em regular} gauges, which alter the form of the
solution without changing its essential characteristics,
i.e. living the solution in the same sector.

In contrast, by performing a {\em singular} gauge transformation one can change
geometrical, topological and physical properties of the
solution, making a transition to another gauge sector.

In a geometrical sense the solutions of Eqs.\eqref{Einst}
define different 4D pseudo-Riemannian spacetime manifolds
${\cal M}^{(1,3)}\{g_{\mu\nu}(x)\}$.

A subtle point in the GR formalism is that transitions from a given physical
solution to an essentially different one is traditionally dubbed a "change of
coordinates".

From the point of view of gauge theory this is confusing, because the choice of space-time coordinates in GR means simultaneously two
different procedures: 1) choosing a gauge sector in which the solution
lives; and at the same time 2) fixing the (regular) gauge in this sector.

While the procedure 2) solves an inessential local gauge fixing problem: change of the labels of the spacetime points,
the above procedure 1) fixes the essential global characteristics of the solution,
i.e. their physical properties and physical meaning.

One can understand better this peculiarity of GR in the framework
of its modern differential geometrical description as a gauge
theory, using the principal frame bundle $\mathcal{F}\left({\cal
M}^{(1,3)}\{g_{\mu\nu}(x)\}\right)$. (See, for example, the second
edition of the monograph  \cite{Exact}, \cite{GS} and
references therein.).

The change of coordinates on the base
space ${\cal M}^{(1,3)}\{g_{\mu\nu}(x)\}$ induces {\em
automatically} a nontrivial change of the frames on the fibre of
frames.

Singular coordinate transformations can produce a
change of the gauge sector of the solution, because they may
change the topology of the frame bundle, adding new singular
points and(or) singular sub-manifolds or removing some of the
existing ones.

For example, singular transformations can alter the
fundamental group of the base space ${\cal M}^{(1,3)}\{g_{\mu\nu}(x)\}$, the holonomy group of the affine
connection on
$\mathcal{F}\left({\cal M}^{(1,3)}\{g_{\mu\nu}(x)\}\right)$, etc.

Thus, one sees that coordinate changes
in our broad sense may be much more than a pure alteration of the
labels of space-time points.

If one works in the framework of the theory of {\em smooth real}
manifolds, ignoring the analytical properties of the solutions
in the complex domain, it seems that one is allowed to change the gauge
without altering the physical problem in the {\em real}
domain, i.e. without change of the boundary conditions,
as well as without introduction of new singular points in the real domain,
or change of the character of the existing ones in this domain.

Such special type of {\em regular}
gauge transformations in GR describes the diffeomorphisms of
the {\em real} manifold ${\cal M}^{(1,3)}\{g_{\mu\nu}(x)\}$.

This manifold is already fixed by the initial choice of the gauge.
Hence, the {\em real} manifold ${\cal M}^{(1,3)}\{g_{\mu\nu}(x)\}$
is actually described by a class of equivalent gauges, which correspond
to all diffeomorphisms of this manifold and are related with regular gauge
transformations in the real domain.

From the physical point of view one must make difference between changes of variables in the 4D spacetime
manifold ${\cal M}^{(1,3)}\{g_{\mu\nu}(x)\}$ and changes of variables in
the 3D space ${\cal M}^{(3)}\{-g_{mn}({\bf r })\}$, $m,n=1,2,3$.

While in the last 3D case we indeed have a change of coordinates,
which at first glance looks like a formal mathematical procedure,
more general changes of coordinates in the first 4D case include changes
of the reference frame (i.e. the reference physical body), which is certainly a physical
procedure.

Further on we will consider only changes of coordinates in the Riemannian 3D manifold
${\cal M}^{(3)}\{-g_{mn}({\bf r })\}$ in a fixed reference frame.

The transitions between some specific real 3D manifolds
${\cal M}^{(3)}\{-g_{mn}({\bf r })\}$, which are not diffeomorphic, can be
produced by the use of proper {\em singular} gauge
transformations.

If we start from an everywhere smooth in the real domain manifold,
after a singular transformation we will have a new 3D manifold with
some different singularities, which possibly lie only in the {\em complex} domain.

The new singularities are to describe physical
phenomena in the problem at hand as the corresponding properties of its solutions.

The inclusion of such singular
transformations in our consideration yields the necessity to introduce
gauge transformations {\em in a broad sense}. They are
excluded from present-day standard considerations by the commonly
used assumption that in GR one has to allow {\em only}
diffeomorphic mappings of real manifolds.

Similar singular transformations are well known in gauge theories
of other fundamental physical interactions: electromagnetic
interactions, electroweak interactions, chromodynamics.

For example, in the gauge theories singular gauge
transformations describe transitions between solutions in
topologically different gauge sectors.
Singular gauge transformations are used in the theory of
Dirac monopole, vortex solutions, t'Hooft-Poliakov monopoles,
Yang-Mills instantons, etc.
See, for example, \cite{Rubakov, GS} and references therein.

The simplest example is the singular gauge transformation of the
3D vector potential: ${\bf A}({\bf r}) \to {\bf A}({\bf r})+\nabla
\varphi({\bf r})$ in electrodynamics, defined in the Cartesian
coordinates $\{x,y,z\}$ by the singular gauge function
$\varphi=\alpha \arctan(y/x)$ ($\alpha=const$).

Suppose that before the transformation we had a 3D flat Euclidean space
${\cal M}^{(3)}\{-g_{mn}({\bf r})=\delta_{mn}\}={\mathcal
E}^{(3)}\{\delta_{mn}\}$. Then this singular gauge transformation
removes the whole axis $OZ$ out of the 3D space ${\mathcal
E}^{(3)}\{\delta_{\mu\nu}\}$, changing the topology of this part of
the base space. As a result the quantity $\oint_Cd{\bf r} {\bf
A}({\bf r})$, which is gauge invariant under regular gauge
transformations, now changes its value to $\oint_Cd{\bf r} {\bf
A}({\bf r})+2\pi N \alpha$, where $N$ is the winding number of the
cycle $C$ around the axis $OZ$.

Under such singular gauge
transformation the solution of some initial physical problem will
be transformed into a solution of a completely different physical problem.

At present, the role of singular gauge transformations in the above
physical theories is well understood.

In contrast, we still have no systematic study of the classes
of physically, geometrically and topologically different solutions
in GR created by singular gauge transformations even in the simple
case of static spherically symmetric space-times with only one point
singularity, although the first solution of this type was discovered
first by Schwarzschild more than one century ago,
i.e. as early as 1916 \cite{Schwarzschild}.

Moreover, in GR at present there is still no clear understanding of both the
above gauge problem and its physical consequences.

Here we present some initial steps toward the clarification of the role
of different GR gauges in a broad sense for spherically symmetric
static space-times with point singularity at the center of symmetry
and vacuum outside this singularity.

\section{GR Static Spherically Symmetric Space-Times with a Single Point Source of
Gravity at rest}

\subsection{The Three Dimentional Riemannian Geometry of the Point Particle Problem}

The single point particle with proper rest mass $M_0$
can be treated as a 3D entity. Its proper frame of reference is
most suitable for description of the {\em static} space-time with this
single particle in it.

We prefer to reduce the problem of single point source
of gravity in GR to a 1D mathematical problem, considering the
dependence of the corresponding functions on the only essential
variable -- the radial variable $r$. This can be achieved in
the following way.

The spherical symmetry of the 3D space reflects adequately
the non-rotating-point character of the source of gravity.

The spherically symmetric 3D Riemannian space
${\cal M}^{(3)}\{-g_{mn}({\bf r })\} \subset {\cal
M}^{(1,3)}\{g_{\mu\nu}(x)\}$, $m,n=1,2,3$,
can be described using the standard spherical coordinates
$r\in[0,\infty),\,\theta\in[0,\pi),\phi\in[0,2\pi)$
in some auxiliary Euclidean space ${\cal E}^{(3)}$\footnote{
The natural idea to use such auxiliary Euclidean space in
the problem under consideration  can be found already in \cite{Schwarzschild},
but Schwarzschild applied different coordinates in it.}.
Then, as usual
$$ {\bf r}=\{x^1,x^2,x^3\}=\{r\sin\theta\cos\phi,r\sin\theta\sin\phi,r\cos\theta\}$$
and an Euclidean squired 3D infinitesimal distance
$$dl^2=dr^2+r^2 d\Omega^2.$$
Here and further on $d\Omega^2=d\theta^2+\sin^2\theta\,d\phi^2$ is the squired interval on the 2D unit sphere in the 3D Euclidiean space.

The role of GR is that the Einstein Eqs.\eqref{Einst} transform the Euclidean space
${\cal E}^{(3)}\{-\eta_{mn}(x^1,x^2,x^3)\} $ to some curved Riemannian one
${\cal M}^{(3)}\{-g_{mn}((x^1,x^2,x^3))$ with a squired 3D infinitesimal distance

\ben
dl^2=- g_{rr}(r)\,dr^2+\rho(r)^2 d\Omega^2
\la{dl0}
\een
with beforehand unknown functions $g_{rr}(r)<0$ and $\rho(r)>0$, which have to be obtained from the Einstein equations \eqref{Einst}.

The physical and geometrical meaning of the radial coordinate $r$ in space ${\cal M}^{(3)}\{-g_{mn}({\bf r })\}$
is not defined by the spherical symmetry of the problem and is
unknown \textit{a priori} \cite{Eddington}.

The only clear thing
is that its value $r=0$ corresponds by construction to the center of the symmetry,
where one must place the physical source of the gravitational
field\footnote{In the present article, we assume that in the Universe
other sources of the gravitational field outside the center of
symmetry do not exist.}.

This fact permits us to identify the  Euclidean space
${\cal E}^{(3)}$ with the tangent space above the place of the point source of gravity, which clarifies
its geometric meaning from a point of view of modern differential geometry.

Now the physical meaning of the space ${\cal E}^{(3)}\{-\eta_{mn}\}$ becomes clear.
This Euclidean space is precisely the tangent 3D space to
${\cal M}^{(3)}\{-g_{mn}\}$ at the point $r=0$ where we are performing all our measurements with our laboratory equipment
staying in the rest reference frame of the point particle.

The space ${\cal E}^{(3)}\{-\eta_{mn}\}$ is a subspace of the tangent space to the
4D basic manifold ${\cal M}^{(1,3)}\{g_{\mu\nu}(x)\}$
above the proper point with $t$ -- arbitrary fixed in the rest frame of the point particle.

In this 4D tangent space the special relativity is locally valid, together with all established physical laws.

GR permits us to relate the measurements in this reference system with measurements
in other ones above other points in the base space ${\cal M}^{(1,3)}\{g_{\mu\nu}\}$.

The radial 3D geometrical distance from the point source at the center of the symmetry is
\ben
l(r)=\int_0^r \sqrt{-g_{rr}(r)} dr
\la{ldistance}
\een

The quantity $\rho$ in Eq.\eqref{dl0} has a clear
geometrical and physical meaning: It is well known that $\rho$
defines the area
\ben
A=4\pi\rho^2
\la{Area}
\een
of a centered at $r=0$ sphere
with a {\em luminosity} radius $\rho$ and the length of a big circle on it
\ben
l_\rho=2\pi\rho.
\la{lcircle}
\een
Relations \eqref{Area} and \eqref{lcircle} only resemble the Euclidean geometry ones
since the variable $\rho$ does not measure geometrical distances in the curved space ${\cal M}^{(3)}\{-g_{mn}\}$.

One can refer to the quantity  $\rho$ as an {\em ``area
radius''},
or as an optical
{\em "luminosity distance``}, because the luminosity of
distant physical objects is reciprocal to $A$.

This consideration shows that the area radius $\rho$ is an invariant quantity
and its value does not depend on the choice of coordinates in the 3D Riemannian space ${\cal M}^{(3)}\{-g_{mn}\}$.

Most probably this property stimulated Hilbert to choose the luminosity distance $\rho$
as a radial coordinate in the Schwarzschild problem \cite{Hilbert}.

We shall also mention that in 3D geometry with metric \eqref{dl0}
the 3D volume of a ball with r-radius $r$ and the center at the origin is
\ben
V^{(3)}(r)=4\pi \int_0^r \sqrt{-g_{rr}(r)}\, \rho(r) r dr.
\la{3DV}
\een

\subsection{The Spacetime Pseudo-Riemannian Geometry and Solutions of the Einstein Equations Outside the Point Source}

Now going to the geometry of the 4D pseudo Riemannian manifold ${\cal M}^{(1,3)}\{g_{\mu\nu}(x)\}$ of signature $\{+,-,-,-\}$
we have to stress that there exists unambiguous choice of a global time $t$ due to the
requirement to use a static metric when we are in the rest frame of the point particle placed at the geometrical point $r=0$.

This yields a familiar form of the space-time interval \cite{ComJann}:
\ben
ds^2=g_{tt}(r)\,dt^2+g_{rr}(r)\,dr^2-\rho(r)^2d\Omega^2 \la{ds0}
\een
with unknown functions $g_{tt}(r)>0,\,g_{rr}(r)<0,\,\rho(r)$, which have to be obtained from equations of the physical
theory of gravity.

Note that in the frame of free falling clocks,
if $\rho_{fixed}$ is some arbitrary fixed value of the luminosity distance, at which the clocks start to fall with zero velocity,
the expression
$$\left(\rho_2-\rho_1\right)/\left( 1-2M/\rho_{fixed}\right)^{1/2}$$
measures the 3D geometrical distance between the geometrical points
$2$ and $1$ on a radial geodesic line \cite{GH}.

Nevertheless, even in this frame the absolute value of the variable
$\rho$ remains not fixed by the 3D distance measurements.

In the static spherically symmetric case
the choice of spherical coordinates and static metric
dictates the form of three of the gauge fixing
coefficients (\ref{Gammas}):\,
$\bar\Gamma_t\!=\!0,\,\,\bar\Gamma_\theta\!=
-\!\cot\theta,\,\,\bar\Gamma_\phi\!=\!0$,
but the form of the quantity
$\bar\Gamma_r\!=\left(\!
\ln\left({\sqrt{-g_{rr}}\,{\bar\rho}^2}\over{\sqrt{\,\,g_{tt}}\,\,\rho^2}\right)\!\right)^\prime$,
or, equivalently, the function $\rho(r)$
are still not fixed. Here and further on, the prime denotes
differentiation with respect to the variable $r$.

We refer to the
freedom of choice of the function $\rho(r)$ as "{\em rho-gauge
freedom}", and to the choice of the $\rho(r)$ function as
"{\em rho-gauge fixing}" in a {\em broad} sense.

In the present article, we will not use more general gauge
transformations in a broader sense than the rho-gauge ones.

In our 1D approach to the problem at hand all possible
mathematical complications, due to the use of such a wide class
of transformations, can be easily controlled.

The overall action for the aggregate of a point particle and its
gravitational field is ${\cal A}_{tot}\!=\!{\cal A}_{GR}+{\cal
A}_{M_{0}}$. Neglecting the surface terms, one can represent the
Hilbert-Einstein action
\ben
{\cal A}_{GR}\!=\!-{1\over{16\pi
G_n}}\int\!d^4x\sqrt{|{}^{(4)}g|}R
\la{GRAction}
\een
and the mechanical action
\ben
{\cal
A}_{M_{0}}\!=\!-M_0\int\!ds
\la{PointAction}
\een
of the point source with proper mass
$M_0$ as integrals with respect to the time $t$ and the radial
variable $r$ of the following Lagrangian densities:
\ben
{\cal L}_{GR}={\frac {1} {2 G_N}}\!\left({
{2\rho\rho^\prime\left(\sqrt{g_{tt}}\right)^\prime\!+\!
\left(\rho^\prime\right)^2\sqrt{g_{tt}} } \over {\sqrt{\!-g_{rr}}}
}\!+\!\sqrt{g_{tt}}\sqrt{\!-g_{rr}}\right),\nonumber\\ {\cal
L}_{M_{0}}= - M_0\sqrt{g_{tt}}\delta(r).\hskip 4.85truecm
\la{LGR}
\een
Here $G_N$ is the Newton gravitational constant,
$\delta(r)$ is the 1D Dirac function \cite{Gelfand}. (In the
units $c\!=\!1$.)

As a result of the rho-gauge freedom, the field variable
$\sqrt{\!-g_{rr}}$ is not a true dynamical variable but rather
plays the role of a Lagrange multiplier, which is needed in the
description of the constrained dynamics.

This auxiliary variable
enters into the Lagrangian ${\cal L}_{GR}$ in {\em a nonlinear} manner,
in contrast to the case of the standard Lagrange multipliers.

The corresponding Euler-Lagrange equations read:
\ben
\left({{2\rho\rho^\prime}\over{\sqrt{\!-g_{rr}}}}\right)^\prime
-{{{\rho^\prime}^2}\over{\sqrt{\!-g_{rr}}}}-\sqrt{\!-g_{rr}}+2 G_N
M_0 \delta(r)=0,
\nonumber\\
\left({{\left(\rho\sqrt{\!g_{tt}}\right)^\prime}\over{\sqrt{\!-g_{rr}}}}
\right)^\prime
-{{\rho^\prime\left(\sqrt{\!g_{tt}}\right)^\prime}\over{\sqrt{\!-g_{rr}}}}=0,
\nonumber\\
{{2\rho\rho^\prime\left(\sqrt{g_{tt}}\right)^\prime\!+\!
\left(\rho^\prime\right)^2\sqrt{g_{tt}} } \over {\sqrt{\!-g_{rr}}}
}\!-\!\sqrt{g_{tt}}\sqrt{\!-g_{rr}}\stackrel{w}{=}0
\la{EL}
\een
where the symbol "\,$\stackrel{w}{=}$\," denotes a weak equality
in the sense of the theory of constrained dynamical systems.

If one ignores the point source of the gravitational field, thus
considering only the domain $r>0$ where $ \delta(r)\equiv 0$, one
obtains the standard {\em vacuum} solution of this system \cite{ComJann}:
\ben
g_{tt}(r)=1-{{\rho_G}/{\rho(r)}},\,\,\, g_{tt}(r)\, g_{rr}(r)=
-\left(\rho(r)^\prime\right)^2,
\la{ss}
\een where $\rho_G=2G_N M$
is the Schwarzschild luminosity radius, $M$ is the gravitational
(Keplerian) mass of the source, and $\rho(r)$ is
an arbitrary ${\cal C}^1$ function.

\section{Some Examples of Different Radial Gauges }

In the general case of metric \eqref{ds0} with arbitrary function $\rho(r)$ one finds the relation $\bar\Gamma_r=-\bar\varphi^\prime$.  On the solutions of vacuum Einstein equations
\ben
\bar\varphi(r)=\ln\left({\frac{\bar\rho^2}{\rho\left(\rho-\rho_G\right)}}{\frac{d\rho}{dr}}\right).
\la{bar_phi_rho}
\een

\subsection{Examples of Radial Gauges for static spherically symmetric solutions of the Einstein equations}
In the literature one can find different choices of the function
$\rho(r)$ for the Schwarzschild problem.

In this Section, we pay attention only to some of the mathematical achievements
of the cited articles and ignore their incorrect interpretations,
due to the lack in understanding of the gauge dependence of the solutions of the Einstein equations \eqref{Einst}.

1. The original Schwarzschild gauge \cite{Schwarzschild}:
$$\rho(r)=\left( r^3+\rho^3_G\right)^{1/3}.$$

It produces
$$\bar\varphi(r)=\ln\left({\frac{\bar\rho^2 r^2}{\left(r^3+\rho_G^3\right)\left(\left(r^3+\rho_G^3\right)^{1/3}-\rho_G\right)}}\right) $$

2. The Hilbert gauge \cite{Hilbert}:
$$\rho(r)=r.$$
It gives
$$\bar\varphi(r)=\ln \left( {\frac {\bar\rho^2}{r\left(r-\rho_G\right)}}\right). $$

This simple
choice of the function $\rho(r)$ is often related incorrectly with
the original Schwarzschild article \cite{Schwarzschild}
(see for example \cite{AL}.).

In this case the coordinate $r$ coincides with the
luminosity distance $\rho$ and the physical domain $r\in
[0,\infty)$ contains an event horizon at $\rho_{{}_H}=\rho_G$.

This unusual circumstance forces one to develop a nontrivial
theory of BH for the Hilbert gauge -- see for example
\cite{books,FN} and references therein.

3. The Droste gauge \cite{Droste}:

The function $\rho(r)$ is given implicitly by the relations
$$\rho/\rho_G=\cosh^2\psi\geq 1,\,\,\,\,
r/\rho_G=\psi+\sinh\psi\cosh\psi.$$

Now
$$\bar\varphi(r)=\ln \left({\frac{\bar\rho^2/\rho_G^2}{\sinh(\psi)\cosh(\psi)^3}}\right).$$

For this solution the variable $r$ has a clear geometrical meaning: it
measures the 3D-radial distance $l = r$ (See Eq.\eqref{ldistance}.)
to the center of the spherical symmetry.

4. The Weyl gauge \cite{Weyl}:
$$\rho(r)={{\rho_G}\over
4}\left(\sqrt{4 r/{\rho_G}}+\sqrt{{\rho_G}/
4 r}\right)^2\!\geq\!\rho_G,$$
$$\bar\varphi(r)=\ln \left( {\frac {\bar\rho^2}{\left( r-\rho_G/4 \right)\left( r + \rho_G/4 \right)}}\right). $$

In this gauge the 3D-space ${\cal M}^{(3)}\{-g_{mn}({\bf r })\}$
becomes conformally flat. The coordinate $r$ is the
radial variable in the corresponding Euclidean 3D space.

5. The Einstein-Rosen gauge \cite{Einstein}:

In the original article
\cite{Einstein} the variable $u^2=\rho-\rho_G$ was used. To
have a proper dimension, we replace it by $r=u^2\geq 0$. Hence,
$$ \rho(r)=r+\rho_G\geq \rho_G\,\,\, \text{and} \,\,\,
\bar\varphi(r)=\ln \left( {\frac {\bar\rho^2}{r\left( r + \rho_G \right)}}\right). $$

6. The isotropic t-x gauge, defined according to the formula
$$x=\rho+\rho_G\ln\left({{\rho}\over{\rho_G}}-1\right),\,\,\,
\rho\geq\rho_G.$$

The variable $x\in (-\infty,\infty)$ is called the tortoise coordinate \cite{books}.  For reader's convenience we keep the standard notation $x$, instead of introducing notation $r$.

The inverse relation is described by the Lambert W function in the form
$$ \rho(x)=\rho_G(W+1),\quad W=\text{LambertW}\left(\exp\left({\frac x {\rho_G}}-1\right)\right). $$

Then, only the combination $(dt^2\!-\!dx^2)$
appears in the 4D-interval $ds^2$ and
$$ \bar\varphi(x)=2\ln\left({\frac {\bar\rho/\rho_G}{W+1}}\right). $$

7. Several different rho-gauges  in GR were introduced by Pugachev and Gun'ko
and, independently, by Menzel, see \cite{ComJann}.

We will not consider in detail them and restrict ourselves only to mentioning the idea of using the same additional condition
in the flat Minkowski space ${\cal E}^{(1,3)}\{\eta_{\mu\nu}(x)\}$ and in the curved one ${\cal M}^{(1,3)}\{g_{\mu\nu}(x)\}$.

This approach defines the choice of the coordinates in these two different sapcetimes in a coherent way.

Using a zero tensor of affine deformation
$$D^\lambda_{\mu\nu}=
\Gamma^\lambda_{\mu\nu}\left( g_{\alpha\beta}(x),\partial g_{\alpha\beta}(x) \right)-
\Gamma^\lambda_{\mu\nu}\left( \eta_{\alpha\beta}(x),\partial \eta_{\alpha\beta}(x) \right) = 0 $$
in the spherical coordinates $x$, Pugachev and Gun'ko have derived once more the original Schwarzschild solution,
see the above point 1.

Note that the relation $ D^\lambda_{\mu\nu}=0$ is not a gauge condition of the type of Eq. \eqref{Gammas}.

In two successive papers (See in \cite{ComJann}) Kohler introduced a new gauge condition
\ben
\nabla_\alpha^{flat} \nabla_\beta^{flat}\left( g^{\alpha\gamma}g_{\sigma\rho} \nabla_\gamma^{flat}
\left(\sqrt{{\frac {g(x)} {\eta(x)}}}\,g^{\rho\beta} \right) \right)=0.
\la{Kohler_gauge}
\een
and studied some of its applications.

Here $\nabla_\alpha^{flat}$ is the covariant derivative with respect to the flat spacetime metric $\eta_{\alpha\beta}(x)$ and
$\eta(x)=\det(\eta_{\alpha\beta}(x))$, $g(x)=\det(g_{\alpha\beta}(x))$.

8. The coherent gauge:

Let us apply the idea of coherent gauge directly to the quantities $\bar\Gamma_r$ using the gauge condition
$$\bar\Gamma_r\!=\!{2\over r}  \quad \Rightarrow \quad \bar\varphi(r)= 2\ln {\frac {\bar\rho}{r}}$$
in the curved spacetime ${\cal M}^{(1,3)}\{g_{\mu\nu}(x)\}$.
It is identical to the rho-gauge fixing of spherical coordinates in the Minkowski
spacetime ${\cal E}^{(1,3)}\{\eta_{\mu\nu}(x)\}$.

Thus, the spherical coordinates $t, r,\theta,\phi$ in the curved spacetime ${\cal M}^{(1,3)}\{g_{\mu\nu}(x)\}$
are fixed in a complete {\em coherent} way with the flat spacetime spherical ones.

Then we obtain a novel form of the 4D interval for a point source
\ben
ds^2 = e^{2\varphi_{\!{}_N}({\bf
r})}\left(\!dt^2\!-\!{{dr^2}\over{N(r)^4}}\!\right)\!-\!{{r^2}\over{N(r)^2}}d\Omega^2,
\la{ds2PGM}
\een
where
\ben
N(r)\!=\!\left(2\varphi_{\!{}_N}\right)^{-1}\left(e^{2\varphi_{\!{}_N}}-1\right)
\la{NPGM}
\een
$N(r)\!\sim\!1\!+\!{\cal O}({{\rho_G}\over r})$ --
for $r\!\to\!\infty$ and $N(r)\!\sim\!{{r}\over \rho_G}$ -- for
$r\to +0$. In this specific gauge
\ben
\rho(r)= r/N(r)={{\rho_G} \over {1- e^{2\varphi_{\!{}_N}(r)}}}
\la{rhoPGN}
\een

and $\rho(r)\!\sim\!r$ -- for $r\!\to\!\infty$, $\rho(r)\!\to\!\rho_G$
-- for $r\to+0$.

Let us consider in more detail the properties of this not well-known exact solution to Eqs.\eqref{Einst}.

The most remarkable property of this flat spacetime gauge is the role of the classical Newton
gravitational potential
\ben
\varphi_{\!{}_N} ({\bf r})=-{{G_N M}\over r}
\la{phiNewton}
\een
in the above  exact GR solution. As usual, the component
$$g_{tt}=e^{2\varphi_{\!{}_N} ({\bf r})}\sim
1+2\varphi_{\!{}_N}({\bf r})+{\cal O}(\varphi_{\!{}_N}({\bf r}))
\, \text{for}\, r\to \infty$$

Hence, the new form of the metric is asymptotically flat for $r\to\infty$.

For the exact solution at hand the center  $r=0$ is an {\em essential} singular point
of the function $g_{tt}$. This strong singularity is the left boundary of the physical intervals $r \in (0,\infty)$ and $\rho \in (0,\infty)$.

For $r\to+0$ the limit of the squared 4D-interval is
\ben
ds^2 \to \rho_g^2 d\Omega^2.
\la{limds2PGM}
\een
As a result, the area of the sphere centered at $r=0$ has a {\em finite} limit
\ben
\lim_{r\to +0}A(r)=4\pi \rho_G^2.
\la{linAPGM}
\een

When one approaches the center $r=0$ from the right,
the 4D volume and the 3D volume approach zero because
$$\lim_{r\to +0}\sqrt{|{}^{(4)}\!g|}=\lim_{r\to +0}\left(e^{2\varphi_{\!{}_N}({\bf r})}N(r)^{-4} r^2\sin\theta\right)  = 0,$$
and
$$\lim_{r\to +0}\sqrt{|{}^{(3)}\!g|}=\lim_{r\to +0}\left(e^{\varphi_{\!{}_N}({\bf r})}N(r)^{-4} r^2\sin\theta\right)  = 0.$$

The only functionally independent algebraic invariant, the Kretschman one
\ben
dI_0^2={1\over {48}} R_{\mu\nu\lambda\kappa}
R^{\mu\nu\lambda\kappa}={1\over{4 \rho_g^4}}\left(1\!-\!e^{2\varphi_{\!{}_N}({\bf r})}\right)^6={{\rho_g^2} \over {4\rho^6}},
\la{dI02PGM}
\een
is finite everywhere, including the center $r\to +0$.

The same is true for the simplest differential invariant
\ben
dI_1^2\!=\!{1\over {720}} \nabla_\lambda R_{\mu\nu\lambda\kappa} \nabla^\lambda R^{\mu\nu\lambda\kappa}\!
=\!-{ {e^{2\varphi_{\!{}_N}}\left(1-e^{2\varphi_{\!{}_N}({\bf r})}\right)^{8} }\over{4 \rho_g^6}}\,\,\,
\la{dI12PGM}
\een
with a zero limit when approaching the center $r\to +0$, as well as for higher order differential invariants of the type
\ben
dI^2_n\sim\nabla_{\lambda_1} \nabla_{\lambda_2}\dots \nabla_{\lambda_n}
R_{\mu\nu \lambda\kappa} \nabla^{\lambda_1} \nabla^{\lambda_2}\dots \nabla^{\lambda_n}  R^{\mu\nu\lambda\kappa}\,
\la{dIn2PGM}
\een
which have a much more complicated explicit form.

Nevertheless, for odd numbers $n$ the invariants $dI^2_n\to 0$ when $r\to +0$.
For even $n$ the invariants $dI^2_n$ have finite values in this limit.
We directly checked this fact for $n=0,1,2,\dots,9$.

At this point one has to stress two facts:

i) For static spherically symmetric solutions of the Einstein equations \eqref{Einst} in the domain $r>0$
the event horizon $\rho_{{}_H}=\rho_G$ exists
under the Hilbert choice of the function $\rho(r)\equiv r$
but not for all other gauges discussed above.

This demonstrates that the existence of the mathematical Schwarzschild BH strongly depends on the
choice of the rho-gauge in a {\em broad} sense.

ii) The choice of the function $\rho(r)$ can change drastically
the character of the singularity at the place of the point source
of the gravitational field in GR.

\subsection{Some Further Notes and Problems}

New interesting gauge conditions for the radial variable $r$ and
corresponding gauge transformations were introduced and
investigated in \cite {Bel1}. In more recent articles \cite{Bel2} it
was shown that using a nonstandard $\rho$-gauge for the
application of GR in  stelar physics, one can describe ECOs
with {\em arbitrary large} mass, density and size. In
particular, one is able to shift the value of the
Openheimer-Volkoff limiting mass $0.7 M_\odot$ of neutron stars to
an essentially larger one: $3.8 M_\odot$.

This may shed a new light on these
astrophysical problems and once more shows that the choice of
rho-gauge (i.e. the choice of the coordinate $r$) can have
real physical consequences \cite{Fiziev2004}.

Thus, we see that by analogy with the classical electrodynamics and
non-abelian gauge theories of general type, in GR  we must use
a more refined terminology and the corresponding mathematical
constructions.

We already call the choice of the function
$\rho(r)$ "a rho-gauge fixing in a broad sense".

Now we see that different functions $\rho(r)$ may describe {\em different}
physical solutions of the Einstein equations (\ref{Einst}) with the
{\em  same} spherical symmetry in the presence of only one mater
point at the center of symmetry.

As we have seen, the mathematical
properties of the singular point may depend on the choice of the
rho-gauge in a broad sense. Now the problem is to clarify the
physical meaning of the singular points with different
mathematical characteristics.

Some of the static spherically symmetric solutions with a
singularity at the center can be related using regular gauge
transformations.

For example, choosing the Hilbert  gauge we
obtain a black hole solution of the {\em vacuum} Einstein
equations \eqref{Einst}. Then we can perform a {\em regular} gauge
transformation to harmonic coordinates, defined by the relations
$\bar\Gamma_t=0$, $\bar\Gamma_{x^1}=0$, $\bar\Gamma_{x^2}=0$,
$\bar\Gamma_{x^3}=0$, preserving the existence of the event
horizon, i.e. staying in the same gauge sector.

In the next sections, we will find geometrical criteria for answering the
question: When do different representations of
static spherically symmetric solutions with
point singularity describe diffeomorphic spacetimes ?

The definition of the regular gauge transformations allows
transformations that do not change the number and the character of
singular points of the solution and the
behavior of the solution at the boundary of the physical domain
but can place these points and the boundary in new positions.

\section{General Form of the Gravitational Field Equations in the Hilbert Gauge
and Their Solutions}

According to Lichnerowicz \cite{Lichnerowicz, YB, Exact}, the
physical spacetimes ${\cal M}^{(1,3)}\{g_{\mu\nu}(x)\}$ of
general type must be smooth manifolds at least of class $C^2$ and
the metric coefficients $g_{\mu\nu}(x)$ have to be at least of
class $C^3$, i.e. at least three times continuously differentiable
functions of the coordinates $x$. When one considers $g_{tt}(r)$,
$g_{rr}(r)$ and $\rho(r)$ not as distributions but as usual
functions of this class, one is allowed to use the rules of the
analysis of functions of one variable $r$. Especially, one can
multiply these functions and their derivatives of the
corresponding order, raise them at various powers, define
functions like $\log$, $\exp$ and other mathematical functions of
$g_{tt}(r)$, $g_{rr}(r)$, $\rho(r)$ and the corresponding derivatives.
In general, these operations are forbidden for distributions
\cite{Gelfand}.

In addition, making use of the standard rules for operating with
the Dirac $\delta$-function \cite{Gelfand} and accepting (for
simplicity) the following assumptions: 1) the function $\rho(r)$
is a monotonic function of the {\em real} variable $r$; 2)
$\rho(0)=\rho_{min}$ and the equation $\rho(r)=\rho_{min}$ has only one
{\em real} solution: $r=0$; one obtains the following alternative
form of Eq.(\ref{EL}):
\ben
\left(\!\sqrt{\!-g_{\rho\rho}}\!-\!{1
\over{\sqrt{\!-g_{\rho\rho}}}}\!\right)\!{d\over{d
\ln\rho}}\!\ln\!\left(\! \rho\left(\!{1\over
{\!-g_{\rho\rho}}}\!-\!1\!\right)\!\right)\!=\nonumber\\ 2
\sigma_0 G_N M_0\,\delta(\rho\!-\!\rho_{min}),\nonumber\\ {{d^2\ln
g_{tt}}\over{(d \ln\rho})^2}\!+\!{1\over 2}\left(\!{{d\ln
g_{tt}}\over{d \ln\rho}}\!\right)^2\!\!+\!\left(\!1\!+\!{1\over
2}{{d\ln g_{tt}}\over{d \ln\rho}}\!\right){{d\ln
g_{\rho\rho}}\over{d \ln\rho}}\!=\!0, \nonumber\\ {{d\ln
g_{tt}}\over{d \ln\rho}}+g_{\rho\rho}+1\stackrel{w}{=}0.\,
\la{ELrho}\een

Note that the only remnant of the function $\rho(r)$ in the system,
Eq.(\ref{ELrho}), is the numbers $\rho_{min}$ and
$\sigma_0=\text{sign}(\rho^\prime(0))$, which enter only into the first
equation related to the source of gravity. Here $g_{tt}$ and
$g_{\rho\rho}=g_{rr}/ {(\rho^\prime)}^2$ are considered as
functions of the independent variable $\rho \in [\rho_{min},\infty)$.

The form of  Eq. (\ref{ELrho}) shows why one is tempted to
consider the Hilbert gauge as a preferable one: In this form the
arbitrary function $\rho(r)$ "disappears"
and it seems that one can not pay attention to the problem of the choice of gauge.

This is obviously a wrong conclusion from both mathematical and physical reasons.

One usually ignores the general case of an arbitrary value
$\rho_{min}\!\neq\!0$ accepting, without any justification, the value
$\rho_{min}=0$, which seems to be natural in the Hilbert gauge. Indeed, if
we erroneously consider the luminosity distance as a measure of the {\em real}
geometrical distance to the point source of gravity in the 3D
space, we have to accept the value $\rho_{min}=0$ for the position of
the point source. Otherwise, the $\delta$-function term in Eq.
(\ref{ELrho}) will describe a {\em shell} with the radius $\rho_{min}\neq 0$,
instead of a {\em point} source.

Actually, according to our previous consideration, the point source
has to be described using the function  $\delta(r)$. The physical
source of gravity is placed at the point $r=0$ by definition.
There is no reason to change this initial position of the source,
or the interpretation of the variables in the problem at hand.

To what value of the luminosity distance $\rho_{min}=\rho(0)$ corresponds
the {\em real} position of the point source is not known {\em a
priori}. This depends strongly on the choice of the rho-gauge.

One cannot exclude such nonstandard behavior of the {\em physically
reasonable} rho-gauge function  $\rho(r)$, which leads to some
value $\rho_{min} \neq 0$. Physically, this means that instead of
infinity, the luminosity of the {\em point} source will go to a
{\em finite} value, when the distance to the source goes to zero,
as it happens to the Pugachev-Gun'ko-Menzel solution, see Section IV.A.

This very interesting new possibility appears in curved
space-times due to their unusual geometrical properties. It may
have a great impact for physics and deserves further careful
investigation.

If one accepts the value $\rho_{min}=0$, one has to note
that the Hilbert-gauge geometrical singularity at $\rho=0$ will be {\em space-like,}
not time-like.

Indeed, because of the well-known change of the sighs of the components
$g_{tt}(r)$ and $g_{rr}(r)$ in the interior of BH, i.e., for $\rho\in [0..\rho_g]$
the signature of the metric becomes $\{-,+,-,- \}$ and
the physical meaning of the variables $t$ and $\rho$ in this domain is interchanged:
Now the "interior time" is $\rho$ and interior "radial variable" is $t$.

This is a quite unusual {\em nonphysical} property
for a physical source of gravity.

In this case, there is no reason at all to speak about the "center of spherical symmetry",
as people often do.

Actually, it is well known that in this domain the solution of Eqs.\eqref{Einst} is not static.

Moreover, in the case of the Hilbert gauge the singular point $\rho=0$ physically defines
the future-time-infinity in the BH interior
and does not define at all the center of spherical symmetry.

The choice of  values of the parameters  $\rho_{min}$ and
$\sigma_0$ was discussed in \cite{Abrams}. There, a new argument
in favor of the choice  $\rho_{min}=\rho_G$ and $\sigma_0=1$ was given
for different physical problems with spherical symmetry  using an
analogy with the Newton theory of gravity.

In contrast, in the present
article we shall analyze the physical meaning of  different
values of the parameter $\rho_{min}\geq 0$. They turn out to be related
with the gravitational mass defect.

The solution of the subsystem formed by the last two equations of
the system (\ref{ELrho}) is given by the well-known functions
\ben
g_{tt}(\rho)\!=\!1\!-\!{{\rho_G}/{\rho}},\,\,\,
g_{\rho\rho}(\rho)=-1/g_{tt}(\rho).
\la{HilbertSol}
\een Note that
in this subsystem one of the equations is a field equation but
the other one is a constraint.

However, these functions do not
solve the first of the equations (\ref{ELrho}) for any value of $\rho_{min}$,
if $M_0\neq 0$.

Indeed, for these functions the left hand side of
the first equation equals identically zero and does not have a
$\delta$-function-type of singularity, in contrast to the
right-hand side. Hence, the first field equation remains unsolved, if $M_0\neq 0$.

Thus, we see that:

1) Outside the singular point, i.e. for $\rho>\rho_{min}$, the
Birkhoff theorem is strictly valid and we have the standard Hilbert form \eqref{HilbertSol} of
the solution, when expressed through the luminosity distance
$\rho$.

2) The assumption that $g_{tt}(r)$, $g_{rr}(r)$ and $\rho(r)$ are
usual $C^3$ smooth functions, instead of distributions, yields a
contradiction, if $M_0\neq 0$.

This way one is not able to describe correctly the gravitational field of a massive point
source of gravity in GR.

For this purpose the first derivative with respect to the variable $r$ at least of one
of the metric coefficients $g_{\mu\nu}$ must have a finite jump,
needed to reproduce the Dirac $\delta$-function in the
energy-momentum tensor of the {\em massive} point particle.

3) The widespread form of the Schwarzschild solution in the Hilbert
gauge \eqref{HilbertSol} does not describe a gravitational field
of a massive point source and corresponds to the case of vacuum solution with $M_0=0$.

The Hilbert solution does not belong to any physical gauge sector of the gravitational
field created by a massive point source in GR.

Obviously, this vacuum solution has a pure geometrical-topological nature and may be used in
the attempts to reach a pure geometrical description of "matter
without matter" \cite{Wheeler1955}.

In the standard approach to this solution no proper
mass distribution has ever been introduced, since it is identically zero.

The above consideration confirms the conclusion, which was
reached in \cite{ADM} using the isotropic gauge for the Hilbert solution.

In contrast to \cite{ADM}, in the next Sections we will show how
one can include in GR {\em neutral}
particles with nonzero proper mass $M_0$. The corresponding new solutions
are related to the Hilbert one via singular gauge transformations
and belong to a different gauge sector of GR as a nonlinear gauge theory of gravity.

\section{On the Schwarzschild Singularity}

Usually, the Schwarzschild singularity $\rho=\rho_G$ is erroneously related {\em only} with the singularity of the metric \eqref{ds0}
with components \eqref{HilbertSol}, or with algebraic invariants of the Riemannian tensor.

The presence of such a singularity in metric is coordinate dependent and
can be removed from the metric by proper coordinate transformations, see details in \cite{books}.

The absence of this singularity in the algebraic invariants of the Riemannian tensor does not prove
that it cannot be present in other geometrical  or physical quantities, as we will see in this Section.

\subsection{A simple pure geometrical meaning of the Schwarzschild Singularity}

From Eq.\eqref{dl0} one obtains the coordinate-independent-relation
\ben
{\frac {dl} {d\rho} } = 1 \big/ \sqrt{ 1-{ \rho_G/\rho} }.
\la{dldrho}
\een

It shows that for the Schwarzschild solution the function $l(\rho)$
has infinite derivative at the Schwarzschild luminosity radius $\rho_G$.

Obviously, this is a geometrical singularity. It is invariant and can not be removed by
any coordinate transformation in the spacetime.

This consideration recovers part of the true geometrical meaning of the notion of the Schwarzschild singularity.

Relation \eqref{dldrho} can also be rewritten in the following geometrical
form\footnote{Taking into account Eq.\eqref{Area},  we see
that actually relation \eqref{dA} represents an invariant ordinary
differential equation of first order for
the function $A(l)$ with only one parameter $\rho_G$ in it.}
\ben
{\frac {dA} {dl} } = 8\pi \sqrt{\rho(\rho-\rho_G)}.
\la{dA}
\een

It shows that in the case of the Schwarzschild solution the point $\rho=\rho_G$ is
a stationary point of the area function $A(l)$ and stresses a special character
of the corresponding value of the area $A_G=16\pi G_N^2 M^2$.
This value is well known from its role in the theory of BH.

The above property of the area $A$ justifies the geometrical meaning of the Schwarzschild singularity.

\subsection{The Schwarzschild Singularity in the Regge-Wheeler and Zirilli equations for QNMs of GW}

As stressed in the Introduction, QNMs of GW are a very hot topic in the era of gravitational astronomy.

The first description of these QNMs was given by the Regge-Wheeler and Zirilly radial differential equations \cite{RW,Z,Chandrasekhar1983}
\ben
g_{tt}(\rho){\frac {d}{d\rho}}\left(g_{tt}(\rho){\frac {d\mathcal{R}}{d\rho}}\right)+ \left(\omega^2-V(\rho)\right)\mathcal{R}(\rho)=0,
\la{Requation}
\een
where $\rho\in (0,\infty)$ and for $L=l(l+1)$, $l=2,3,\dots$ the functions
\begin{subequations} \la{potentials:ab}
\ben
V_{RW}(\rho)&=&g_{tt}(\rho)\left( {\tfrac L {\rho^2}}-{\tfrac 3 {\rho^3}}\right), \la{potentials:a}\\
V_{Z}(\rho)&=&g_{tt}(\rho)\,{\tfrac {L(L-2)^2\rho^3+3\left( (L-2)\rho+3/2)\right)^2+9/4}{\rho^3((L-2)\rho+3)^2}}  \la{potentials:b}
\een
\end{subequations}
define the Regge-Wheeler (RW) and Zerilli (Z) potentials $V(\rho)$ in Eq.\eqref{Requation}, respectively.

It is transparent that the Schwarzschild singularity $\rho=\rho_G$ is a regular singularity (a simple pole)
of these second order differential equations of Frobenius type.

This singularity is an additional one
to the other two singularities: the regular singular point $\rho=0$ and the irregular singular point $\rho=\infty$.

Due to the presence of these three unremovable by radial-coordinate-change singular points,
which are the only ones in the whole complex domain of the variable $\rho$,
the exact solutions $\mathcal{R}(\rho)$ of the above basic equations can be represented in terms of the confluent Heun functions,
as shown for the first time in \cite{Fiziev2006, Fiziev2009, Fiziev2011}.

If one wrongly supposed, that the Schwarzschild singularity $\rho=\rho_G$ can be removed
from  Eqs.\eqref{Requation}-\eqref{potentials:ab}, this would lead to a completely different type
of the exact QNM solutions and spectra of GW in GR\footnote{For example,
would these equations somehow had only two singularities: $\rho=0$, and $\rho=\infty$,
their solutions would be represented in terms of the confluent hypergeometric functions,
which solve the Hydrogen problem. The spectrum of the QNMs of GW would be similar
to the spectrum of the Hydrogen. This would be a  certainly wrong result,
both mathematically and physically.}.

Solving the two-singular-points boundary problem for Eqs.\eqref{Requation}-\eqref{potentials:ab}
in the physical domain $\rho \in (\rho_G, \infty)$,  in \cite{Fiziev2006, Fiziev2009, Fiziev2011},
we obtained with very good precision (much higher than the previously published ones)
the frequencies of the BH QNMs: $\omega_{n,l}, \, n=0,\dots, l=2, \dots$ of two types of GW (axial and polar) in GR.

These frequencies are physical quantities and their values are defined
by the existing three singularities in Eqs.\eqref{Requation}-\eqref{potentials:ab}.

One can extract the authentic values of these physical quantities from the observational
data by the LIGO/VIRGO collaboration, see \cite{LOGOOPEN,Barack2018,Testing2019} and references therein.

Of course, the physical properties of the spectrum of QNMs do not depend on
the choice of coordinates in the Riemann space ${\cal M}^{(3)}\{-g_{m n}(x)\}$.

At the same time, the physical spectrum of QNMs depends essentially on the presence of the real Schwarzschild Singularity
$\rho=\rho_G$ in Eqs.\eqref{Requation}-\eqref{potentials:ab},
{\em together} with the two others: $\rho=0$, and $\rho=\infty$,
as well as on the absence of any additional singularities.

We have to remind the reader that the Regge-Wheeler and Zirilli equations are linear approximations
to the Einstein equations \eqref{Einst} for weak GW on the Schwarzschild background.

It is extremely interesting that in domain of QNM ringing the nonlinearity of the Einstein equations is not seen.
For example, in GW from binary BH merger: "Rather than nonlinearities, times around the peak
are dominated by ringdown overtones -- the quasinormal
modes with the fastest decay rates, but also the highest
amplitudes near the waveform peak.", see the very recent paper \cite{Testing2019}, and the references therein.

Since the linear QNM solutions of the problem approximate very well
the numerical solutions in full GR in the domain of QNM ringing of BH,
and give a reasonable description of observational data for GW in this domain \cite{Testing2019},
we can expect that the above conclusions about existing of the unavoidable physical Schwarzschild singularity in the QNM problem take place also for the corresponding exact solution of the Einstein equations \eqref{Einst}.

\subsection{Invariants of the Riemann tensor and the Schwarzschild singularity}

It is well known that the algebraic invariants of the Riemann tensor do not see the Schwarzschild singularity $\rho=\rho_G$.
For metric \eqref{ds0} with components \eqref{HilbertSol} there exists only one functionally independent of them -- the Kretschman invariant \cite{books,CMcL}
\ben
dI_0^2={\frac 1 {48}}R_{\alpha\beta\gamma\delta}R^{\alpha\beta\gamma\delta}={\frac 1 4}{\frac {\rho_G^2}{\rho^6}}.
\la{AlgInv}
\een

Since the pure algebraic invariants of the tensor $R_{\alpha\beta\gamma\delta}$ do not fix completely the geometry,
their consideration is not sufficient to recover all gauge-invariant spacetime properties. For
this purpose one must consider a large enough number of high-order differential invariants of
the Riemann tensor \cite{Cartan}.

As early as in \cite{Karlhede1982}, it was recognized that the simplest differential invariant of metric \eqref{ds0} with components \eqref{HilbertSol}
\ben
dI_1^2={\frac 1 {720}}\nabla_\nu R_{\alpha\beta\gamma\delta}\nabla^\nu R^{\alpha\beta\gamma\delta}=-{\frac 1 4}{\frac {\rho_G^2}{\rho^8}}\left(1-{\frac {\rho_G}{\rho}}\right)
\la{DiffInv}
\een
saw as a zero point the Schwarzschild singularity.

This is a complete coordinate-independent result.

As a byproduct, each of Eqs.\eqref{AlgInv}  or \eqref{DiffInv}
proves once more the invariant geometrical nature of the luminosity radius $\rho$,
now in the sense of 4D spacetime geometry.

As seen, the differential invariant \eqref{DiffInv} changes its sign when one crosses the horizon $\rho=\rho_G$.

Physically this means that one can experimentally fix the crossing of the  Schwarzschild singularity and disproves the wide-spread
statement that "nothing special", i.e. measurable, happens when one crosses the horizon of a BH in classical GR.

For the same result for the Kerr metric see \cite{Fiziev2010} and references therein.

A preliminary, still unpublished study shows that the result can be generalized to
a much wider class of solutions of the Einstein equations \eqref{Einst} with horizons.

The higher order differential invariants
\ben
dI^2_n\sim\nabla_{\lambda_1} \nabla_{\lambda_2}\dots \nabla_{\lambda_n}
R_{\mu\nu \lambda\kappa} \nabla^{\lambda_1} \nabla^{\lambda_2}\dots \nabla^{\lambda_n}  R^{\mu\nu\lambda\kappa}
\la{dI2n}
\een
with odd number $n=1,3,\dots$ also see the horizon $r=\rho_G$ as a zero point.
The differential invariants \eqref{dI2n} with even $n=0,2,\dots$ have a finite values at the   horizon $r=\rho_G$.

The full investigation of this problem is not trivial and at present is not carried out even for the Schwarzschild problem.

Obviously, the well-known and wide-accepted opinion that the Schwarzschild singularity is a pure coordinate one is not correct.

This incorrect statement does not solve all geometrical and physical problems related with the Schwarzschild singularity.
The whole story is still with an open end.

\section{Normal Coordinates for Gravitational Field of a
         Massive Point Particle in General Relativity}

An obstacle for the description of the gravitational field of a point
source at the initial stage of development of GR was the absence
of an adequate mathematical formalism.

Even after the development
of the correct theory of mathematical distributions \cite{Gelfand}
there still exists an opinion that this theory is inapplicable to
GR because of the nonlinear character of the Einstein equations
(\ref{Einst}).

For example, the author of \cite{YB}
emphasizes that "the Einstein equations, being non-linear, are
defined essentially, only within framework of functions. The
functionals, introduced in ... physics and mathematics (Dirac's
$\delta$-function, "weak" solutions of partial differential
equations, distributions of Scwartz) are suitable only for linear
problems, since their product is not, in general, defined."

In a more recent article \cite{GT} the authors have considered
singular lines and surfaces using mathematical distributions.

Those authors have stressed that "there is apparently no viable treatment
of point particles as concentrated sources in GR". See also
\cite{Exact} and references therein.

Here we propose a novel
approach to this problem, based on a specific choice of the field
variables.

Let us represent the metric (\ref{ds0}) of the problem at
hand in a specific form:
\ben
ds^2\!=e^{2\varphi_1}dt^2\!-e^{\!-2\varphi_1+4\varphi_2-2\bar\varphi}dr^2\!-
\bar\rho{\,}^2e^{-2\varphi_1+2\varphi_2}d\Omega^2\!,\,\,\,
\la{nc}
\een
where $\varphi_1(r)$, $\varphi_2(r)$ and $\bar\varphi(r)$ are unknown functions of the
variable $r$. The constant $\bar\rho$ is the unit for
luminosity distance $\rho=\bar\rho\, e^{-\varphi_1+\varphi_2}$.

By ignoring the surface terms in the corresponding integrals, one
obtains the gravitational and point particle actions in the form:
\ben
{\cal A}_{GR}\!&=&\!\!{1\over{2 G_N}}\!\!\int\!\!dt\!\!\!\int\!\!dr
\Bigl(e^{\bar\varphi}\left(\!-(\bar\rho\varphi_1^\prime)^2\!+
\!(\bar\rho\varphi_2^\prime)^2\right)
\!+\!e^{\!-\bar\varphi}e^{2\varphi_2}\Bigr),\nonumber\\ {\cal
A}_{M_0}\!&=&\!-\int\!\!dt\!\!\int\!dr\, M_0\, e^{\varphi_1}\delta(r).
\hskip 3.45truecm
\la{A}
\een

Thus, we see that the field variables
$\varphi_1(r)$, $\varphi_2(r)$ and $\bar\varphi(r)$ play the role
of {\em normal fields' coordinates} in our problem. The field
equations read: \ben  \bar\Delta_r \varphi_1(r)= {\frac
{G_NM_0}{\bar\rho^2}} e^{\varphi_1(r)-\bar\varphi(r)}
\delta(r),\nonumber\\ \bar\Delta_r \varphi_2(r)= {\frac
{1}{\bar\rho^2}}
e^{2\left(\varphi_2(r)-\bar\varphi(r)\right)}\hskip 1.1truecm
\la{FEq}
\een
where
$\bar\Delta_r\!=\!e^{\!-\bar\varphi}{d\over{dr}} \left(
e^{\bar\varphi} {d \over{dr}} \right)$ is related to the radial
part of the 3D-Laplacian,
$\Delta\!=\!-1/\sqrt{|{}^{(3)}\!g|}\partial_i
\left(\sqrt{|{}^{(3)}\!g|}g^{ij}\partial_j\right)\!=
\!-g^{ij}\left(\partial^2_{ij}-\bar\Gamma_i\partial_j\right)\!=\!\Delta_r
+1/\rho^2\Delta_{\theta\phi} $
:\,\,\,$\bar\Delta_r\!=\!-g_{rr}\Delta_r$.

The variation of the total action with respect to the auxiliary
variable $\bar\varphi$ gives the constraint: \ben
e^{\bar\varphi}\left(-(\bar\rho\varphi_1^\prime)^2\!+\!(\bar\rho\varphi_2^\prime)^2
\right)-e^{-\bar\varphi}e^{2\varphi_2}\stackrel{w}{=}0.
\la{constraint}\een

One can have some doubts about the correctness of the above
derivation of the field equations (\ref{FEq}), because here we
use the Weyl trick applying the spherical symmetry directly
to the action functional, not to the Einstein equations
(\ref{Einst}). The correctness of the result of this procedure is
proved in Appendix A.

Therefore, if one prefers,
one can consider the Lagrangian densities (\ref{LGR}), or the actions
(\ref{A}) as auxiliary tools for formulating our
one-dimensional problem  as a variational
problem.  defined by the reduced spherically
symmetric the Einstein equations (\ref{Einst}).

The variational approach makes transparent the role of
various differential equations, which govern the problem.
It leads to a simple application of the theory of constrained dynamical systems
in the problem at hand.

\section{Regular Gauges and General Regular Solutions of the Problem}

The advantage of the above normal fields' coordinates is that when
expressed through them, the field equations (\ref{FEq}) are linear with respect to
the derivatives of the unknown functions $\varphi_{1,2}(r)$.

This circumstance legitimates the correct application of the mathematical
theory of distributions and makes our normal coordinates privileged
field variables.

The choice of the function $\bar\varphi(r)$ fixes the rho-gauge in
normal coordinates. We have to choose this function in a way that
makes  the first of the equations (\ref{FEq}) meaningful. Note
that this non-homogeneous equation is quasi-linear and has a
correct mathematical meaning if, and only if, the condition
$|\varphi_1(0)-\bar\varphi(0)|<\infty$ is satisfied.

Let us consider once more the domain $r>0$. In this domain the
first of the equations (\ref{FEq}) gives $\varphi_1(r)=C_1\int
e^{-\bar\varphi(r)}dr+C_2$ with arbitrary constants $C_{1,2}$.
Suppose that the function $\bar\varphi(r)$ has  asymptotic
$\exp(-\bar\varphi(r))\sim k r^n$  in the limit $r\to +0$ (with
some arbitrary constants $k$ and $n$).

Then one easily
obtains $\varphi_1(r)-\bar\varphi(r)\sim C_1 k r^{n+1}/(n+1)+
n\ln{r}+\ln{k}+C_2$ -- if $n\neq -1$ and
$\varphi_1(r)-\bar\varphi(r)\sim (C_1k-1)\ln{r} +\ln{k}+C_2$ --
for $n=-1$.

Now we see that one can satisfy the condition
$\lim_{r\to 0}|\varphi_1(r)-\bar\varphi(r)|= constant <\infty$ for
arbitrary values of the constants $C_{1,2}$ if, and only if $n=0$.

This means that we must have $\bar\varphi(r)\sim k=const \neq
\pm\infty$ for $r\to 0$. We call such gauges {\em regular
gauges} for the problem at hand. Then $\varphi_1(0)=const\neq
\pm\infty$.

Obviously, the simplest choice of a regular gauge is
$$\bar\varphi(r)\equiv 0.$$
Further on we shall use this {\em basic regular gauge}.

Other regular gauges for the same gauge sector
differ from it by a regular rho-gauge transformation,
which describes a diffeomorphism of the
already fixed by the basic regular gauge manifold ${\cal
M}^{(3)}\{-g_{mn}\}$.

In terms of the metric components
the basic regular gauge condition reads
\ben
\rho^4 g_{tt}+\bar\rho^4g_{rr}=0
\la{regcond_g}
\een
and gives
$$\bar\Gamma_r=0.$$

Under this gauge the field equations (\ref{FEq}) acquire the
simplest quasi-linear form: \ben \varphi_1^{\prime\prime}(r)=
{\frac {G_NM_0}{\bar\rho^2}}\, e^{\varphi_1(0)}\,
\delta(r),\,\,\,\, \varphi_2^{\prime\prime}(r)= {\frac
{1}{\bar\rho^2}}\,e^{2\varphi_2(r)}.\hskip .2truecm \la{FEq0}\een

The constraint (\ref{constraint}) acquires a simple form: \ben
-(\bar\rho\varphi_1^\prime)^2\!+\!(\bar\rho\varphi_2^\prime)^2
-e^{2\varphi_2}\stackrel{w}{=}0. \la{constraint0}\een

As can be easily seen, the basic regular gauge $\bar\varphi(r)\equiv 0$ has
the unique property to split the system of field equations (\ref{FEq}) and
the constraint (\ref{constraint}) into three independent relations.

The new field equations (\ref{FEq0}) have a general solution \ben
\varphi_1(r)\!=\!{{G_N
M_0}\over{\bar\rho^2}}\,e^{\varphi_1(0)}\bigl(\!\Theta(r)\!-\!\Theta(0)\!\bigr)\,r
\!+\!\varphi_1^\prime(0)\,r\!+\!\varphi_1(0),\nonumber\\
\varphi_2(r)\!=\!-\ln\left({1\over{\sqrt{2\varepsilon_2}}}
\sinh\left(\sqrt{2\varepsilon_2}
{{r_\infty-r}\over{\bar\rho}}\right)\right). \hskip 1.2truecm
\la{Gsol}\een

The first expression in Eq.(\ref{Gsol}) represents a distribution
$\varphi_1(r)$. In it $\Theta(r)$ is the Heaviside step function.

Here and further on we use the {\em additional assumption} $\Theta(0)=1$. It
gives a specific regularization of the products, degrees and
functions of the distribution $\Theta(r)$ and makes them definite.

For example: $\big(\Theta(r)\big)^2=\Theta(r)$,
$\big(\Theta(r)\big)^3=\Theta(r)$, \dots,
$\Theta(r)\delta(r)=\delta(r)$,
$f(r\Theta(r))=f(r)\Theta(r)-f(0)\big(\Theta(r)-\Theta(0)\big)$ --
for any function $f(r)$ with convergent Taylor series expansion
around the point $r=0$, and so on.

This is the only simple
regularization of distributions we need in the present article.

The second expression $\varphi_2(r)$ in Eq.(\ref{Gsol}) is a usual
function of the variable $r$. The symbol $r_\infty$ is used as
abbreviation for the constant expression
$$r_\infty=\text{sign}\left(\varphi_2^\prime(0)\right)\bar\rho\,
\,\text{arcsinh}\left(\sqrt{2\varepsilon_2}e^{-\varphi_2(0)}\right)/\sqrt{2\varepsilon_2}.$$

The constants
\ben
\varepsilon_1\!=\!-{1\over
2}\bar\rho^2{\varphi_1^\prime(r)}^2\!+\!{{G_N
M_0}\over{\bar\rho^2}}\varphi_1^{\prime}(0)\,e^{\varphi_1(0)}
\bigl(\!\Theta(r)\!-\!\Theta(0)\!\bigr),\nonumber\\
\varepsilon_2\!=\!{1\over
2}\left(\bar\rho^2{\varphi_2^\prime(r)}^2\!-\!e^{2\varphi_2(r)}\right)
\hskip 3.76truecm \la{epsilon_12}
\een
are the values of the
corresponding first integrals (\ref{epsilon_12}) of the
differential equations (\ref{FEq0}) for a given solution
(\ref{Gsol}).

Then for the regular solutions (\ref{Gsol}) the
condition(\ref{constraint0}) reads:
\ben
\varepsilon_1+\varepsilon_2+{{G_N
M_0}\over{\bar\rho^2}}e^{\varphi_1(0)}
\bigl(\!\Theta(r)\!-\!\Theta(0)\!\bigr)
\stackrel{w}{=}0.
\la{constraint_sol}
\een

An unexpected property of this relation is that it cannot be
satisfied for any value of the variable $r\in (-\infty,\infty)$,
because $\varepsilon_{1,2}$ are constants.

The constraint
(\ref{constraint_sol}) can be satisfied either on the interval
$r\in [0,\infty)$, or on the interval $r\in (-\infty,0)$.

If from physical reasons we choose it to be valid at only one point
$r^*\in [0,\infty)$, this relation will be satisfied on the whole
interval $r\in [0,\infty)$ and this interval will be the physically
admissible real domain of the radial variable.

Thus one can see that our approach gives a unique possibility {\em to derive} the
admissible real domain of the variable $r$ from the dynamical
constraint (\ref{constraint0}), i.e., this dynamical constraint
yields a geometrical constraint on the values of the radial
variable.

As a result, in the physical domain the values of the first
integrals (\ref{epsilon_12}) are related by the standard equation
\ben\varepsilon_{tot}=\varepsilon_1+\varepsilon_2\stackrel{w}{=}0,
\la{epsilon}\een
which reflects the fact that our variation problem is invariant
under re-parametrization of the independent variable $r$.

At the end, as a direct consequence of relation (\ref{epsilon}) one
obtains the inequality $\varepsilon_2=-\varepsilon_1>0$, because
in the real physical domain $r\in [0,\infty)$ we have
$\varepsilon_1=-{1\over
2}\bar\rho^2{\varphi_1^\prime(r)}^2=const<0$.

For the function $\rho_{reg,0}(r)\geq 0$, which corresponds to the
basic regular gauge, one easily obtains
\ben\rho_{reg|_{\bar\varphi=0}}(r)=\rho_G
\left(1-\exp\Big({4{{r-r_{\infty}}\over{\rho_G}}}\Big)\right)^{-1}.
\la{reg_rho}\een

Now one has to impose several additional conditions on the
solutions (\ref{Gsol}):

i) The requirement to have asymptotically flat space-time.

The limit $r\!\to\!r_\infty$ corresponds to the limit
$\rho\!\to\!\infty$.

For solutions (\ref{Gsol}) we have the
property $g_{rr}(r_\infty)/{\rho^\prime}^2(r_\infty)\!=\!1$. The
only nontrivial asymptotic condition is $g_{tt}(r_\infty)=1$. It
gives $\varphi_1^\prime(0)\,r_\infty\!+\!\varphi_1(0)\!=\!0$.

ii) The requirement to have the correct Keplerian mass $M$, as seen by
a distant observer.

Excluding the variable $r>0$ from
$g_{tt}(r)=e^{2\varphi_1(r)}$ and $\rho(r)=\bar\rho
e^{-\varphi_1(r)+\varphi_2(r)}$,  for solutions (\ref{Gsol}) one
obtains $g_{tt}=1+{\text{const}\over{\rho}}$, where
$$\text{const}=2\,\text{sign}(\rho)\,\text{sign}(\varphi_2^\prime(0))\,\bar\rho^2\,\varphi_1^\prime(0)=-2G_N M$$.

iii) The consistency of the previous conditions with the relation
$g_{tt}\,g_{rr}+{\rho^\prime}^2=0$ gives
$$\text{sign}(r_\infty)=\text{sign}(\rho)=sign(\varphi_1^\prime(0))=\text{sign}(\varphi_2^\prime(0))=1.$$

iv) The most suitable choice of the unit $\bar\rho$ is
$$\bar\rho=G_N M =\rho_G/2.$$

As a result, all initial constants
become functions of the two parameters in the problem --
$r_\infty$ and $M$:

$\varphi_1(0)= -{{r_\infty}\over{G_N M}}$,

$\varphi_2(0)= -\ln\left(\sinh\left({{r_\infty}\over{G_N
M}}\right)\right)$,

$\varphi_1^\prime(0)= -{{1}\over{G_N M}}$,

$\varphi_2^\prime(0)={{1}\over{G_N
M}}\coth\left({{r_\infty}\over{G_N M}}\right)$.

v) The gravitational defect of the mass of a point particle.

Representing the proper mass $M_0$ of the point source in the form
$$M_0=\int_0^{r_\infty}
M_0\,\delta(r)dr=4\pi\int_0^{r_\infty} \sqrt{-g_{rr}(r)}\,\rho^2(r)\rho^\prime(r)\,\mu(r)dr,$$
one obtains for the mass
distribution of the point particle the expression
$$\mu(r)=M_0\,\delta(r)\Big/\left(4\pi\sqrt{-g_{rr}(r)}\,\rho^2(r)\rho^\prime(r)\right)=M_0\,\delta_g(r),$$
where
$$\delta_g(r)=\delta(r)\Big/\left(4\pi\sqrt{-g_{rr}(r)}\rho^2(r)\rho^\prime(r)\right)=\delta_g(\rho)=inv\{\rho \rightleftarrows r\}$$
is the 1D {\em invariant} under one-to one changes $\rho \rightleftarrows r$ Dirac delta function.

The Keplerian gravitational mass $M$ can be calculated using the components of the massive point particle stress-energy tensor in the Tolman formula
$M= \int d^3{\bf r}\sqrt{\left|{}^{(3)}g\right|}\left(T^0_0-T^1_1-T^2_2-T^2_3\right)$
\cite{books}:
\ben
M = \int_0^{r_\infty} M_0\sqrt{g_{tt}(r)}\,\delta(r)dr= M_0\sqrt{g_{tt}(0)}.
\la{MKepler}
\een

As a result, we reach the important relations:
\ben
g_{tt}(0)&=&e^{2\varphi_1(0)}\!=\!\exp\!\left(\!-2{{r_\infty}\over{G_N M }}\!\right)=\left(\!{{M}\over{M_0}}\!\right)^2\leq1,\nonumber\\
r_\infty&=&G_N M \ln\left({{M_0}\over{M}}\right) \geq 0.
\la{gtt_rho_inf}
\een

Note that due to our convention $\Theta(0)=1$ the component
$g_{tt}(r)$ is a continuous function in the interval $r\in
[0,\infty)$ and $g_{tt}(0)=g_{tt}(+0)$ is a well defined
quantity.

The ratio
\ben
\varrho={{M}\over{M_0}}=\sqrt{g_{tt}(0)}\in [0,1]
\la{varrho}
\een
describes the gravitational mass defect of the point particle as a second
physical parameter in the problem, see the Introduction.

The Keplerian mass $M$ and the ratio $\varrho$ are the two parameters that
define completely the solutions (\ref{Gsol}).

For the initial constants of the problem one obtains:
\ben
\varphi_1(0)&=&\ln\varrho ,\,\,\,\varphi_2(0)=- \ln { {1-\varrho^2}
\over {2\varrho} } ,\nonumber\\ \,\nonumber\\
\varphi_1^\prime(0)&=&
{{1}\over{G_N M}}, \,\,\,\varphi_2^\prime(0)={{1}\over{G_N
M}}\,{{1+\varrho^2}\over{1-\varrho^2}}.
\la{In_constMM}
\een

Thus we arrive at the following form of the solutions (\ref{Gsol}):

\ben
\varphi_1(r)&=&{{r\,\Theta(r)}\over{G_N M}}-\ln(1\!/\!\varrho),
\nonumber\\
\varphi_2(r)\!&=&\!-\ln\left( {1\over
2}\left({1\!/\!\varrho}\,e^{-r/G_N\!M}-
\varrho\,e^{r/G_N\!M}\right)\right)
\la{sol_f}
\een

and the rho-gauge fixing function
\ben
\rho_{reg|_{\bar\varphi=0}}(r)=\rho_G
\left(1-\varrho^2\exp\Big({4{{r}/{\rho_G}}}\Big)\right)^{-1}.
\la{basic_rho_f}
\een

An unexpected feature of this {\em two parametric} variety of
solutions for the gravitational field of a massive point particle is that each
solution must be considered only in the {\em finite} interval
$$r\in [\,0,\,G_N M\ln\left({{1}/{\varrho}}\right)]$$
in the Euclidean 3D space ${\cal E}^{(3)}$ (See Section III.A.)
since we must have $\rho_{reg|_{\bar\varphi=0}}(r) \geq 0$.
Then, we have a monotonic increase of the luminosity distance in the interval
$$ \rho_{reg|_{\bar\varphi=0}}(r) \in [\rho_{min},\infty)$$
for $r$ in the above interval.

It is easy to check that away from the source, i.e., for $r>0$, these
solutions coincide with the solution (\ref{HilbertSol}) in the
Hilbert gauge. Hence, outside the source the solutions
(\ref{sol_f}) acquire the well-known standard form, when
represented using the variable $\rho$.

This means that the solutions (\ref{sol_f}) strictly respect the generalized Birkhoff
theorem. For a massive point source of gravity its proper generalization requires only justification
of the physical domain of the variable $\rho$. It is remarkable
that for the solutions (\ref{sol_f}) the minimal value of the
luminosity distance is
\ben\rho_{min}=\rho_G /(1-\varrho^2)\geq \rho_G.
\la{rho_0}
\een
This changes the Gauss theorem and leads to
important physical consequences for the physical properties of the massive ECOs, see the Introduction.

One such consequence is that in this case one must apply the Birkhoff theorem only in the interval
$\rho\in[\rho_{min},\infty)$, if spherical symmetry takes place.

\section{Regular Gauge Mapping of the Interval $r\in [0,r_\infty]$
onto the interval $r\in[0,\infty]$ of a new auxiliary radial variable $r$}

As we have stressed in the previous section, the solutions
(\ref{sol_f}) for the gravitational field of a point particle must
be considered only in the physical domain of the auxiliary variable $r\in [0,r_\infty]$.

It is not convenient to work with such an unusual radial variable $r$.

Besides, the interpretation of the flat spacetime ${\cal E}^{(3)}$ as a tangent space above the point $r=0$ requires to
include all its points in our consideration.

Remembering that the used until now variable $r$ is a radial coordinate in the initial auxiliary Euclidean space ${\mathcal E}^{(3)}$,
one can easily overcome this problem using an additional regular rho-gauge transformation
\ben
r \mapsto r_\infty{{r/\tilde
r}\over{r/\tilde r+1 }}
\la{rrgt}
\een
with an arbitrary scale $\tilde r$ of the new auxiliary radial variable $r$.

Thus, we replace the initial auxiliary space ${\cal E}^{(3)}$ with another copy of this linear space,
making use of the fractional linear transformation \eqref{rrgt}.

Note that this fractional linear change of the variable $r$ does not
alter the number and the character of the singular points of the
solutions in the whole compactified complex plane $\tilde{\cal C}^{(1)}_r$.

The transformation \eqref{rrgt}, being indeed a  change of the labels in a specific tangent space,
is physically inessential but important for clear understanding of the obtained mathematical results.

To avoid unnecessary notation for the radial coordinate,
in the present article we use the same notation  $r$ for these two different
radial variables.

The transformation (\ref{rrgt})
simply places the point $r=r_\infty$ at the infinity $r=\infty$,
at the same time preserving the initial place of the origin $r=0$.

Now the new variable $r$ varies in the standard interval $r\in
[0,\infty)$.  Assuming $\tilde r>0$ and taking into account that $\Theta(0)=1$, in this interval the regular solutions (\ref{sol_f}) acquire the form
\ben \varphi_1(r)&=&-\ln({1\!/\!\varrho}) \left(1-{{r/\tilde
r}\over{r/\tilde r+1 }}\right),\nonumber\\
\varphi_2(r)&=&-\ln\left( {1\over 2
}\left((1\!/\!\varrho)^{{1}\over{r/\tilde
r+1}}-\varrho^{{1}\over{r/\tilde r+1}}\right)\right);
\la{sol_NewGauge}
\een
and the rho-gauge fixing function reads
\ben\rho_{reg}(r)=\rho_G
\left(1-\varrho^{{{2}\over{r/\tilde r +1 }}}\right)^{-1}.
\la{rho_NewGauge}
\een

The last expression shows that the mathematically admissible
interval of the values of the ratio $\varrho$ is the {\em open}
interval $(0,1)$. This is so, because for $\varrho=0$ and for
$\varrho=1$ we would have impermissible trivial gauge-fixing
functions $\rho_{reg}(r)\!\equiv\!1$ and
$\rho_{reg}(r)\!\equiv\!0$, respectively.

\section{Transformation of the Hilbert solution of the Schwarzschild problem to the solution with
massive point source in GR}

After all, we are ready to describe the singular character of the coordinate
transition from the Hilbert form of the Schwarzschild solution
\eqref{HilbertSol}
to the regular one \eqref{sol_NewGauge}.

To simplify the notation, let us introduce dimensionless variables
$z=\rho/\rho_g$ and $\zeta=r/\tilde r$.
Then Eq. (\ref{rho_NewGauge}) shows that the essential part of the
change of the coordinates is described, in both directions,
by the functions
\ben
z(\zeta)=\left(1-\varrho^{2\over{\zeta+1}}\right)^{-1},\,\,\,
\hbox{and} \nonumber \\
\zeta(z)={{\ln(1/\varrho^2)}\over {\ln z - \ln(z-1)}}-1,
\,\,\,\varrho\in (0,1).
\la{change}
\een

Obviously, the function $z(\zeta)$ is regular at the place of the
point source $\zeta=0$; it has a simple pole at $\zeta=\infty$ and
an essential singular point at $\zeta=-1$. At the same time the
inverse function $\zeta(z)$ has logarithmic branch points at the
Hilbert-gauge center of symmetry  $z=0$ and at the
event horizon $z=1$.

Thus, we see how one produces the Hilbert-gauge
singularities at $\rho=0$ and at $\rho=\rho_{{}_H} \equiv \rho_G$, starting from
a regular solution. The derivative
$$d\zeta/dz={{\ln(1/\varrho^2)}\over{z(z-1)\left(\ln z-\ln(z-1)\right)^2}}$$
approaches infinity at these two points, hence the singular character of the
change in the whole complex domain of the variables.

Now we have a complete description of the change of the coordinates and
singularities of this change in the complex domain of the radial variable.

The restriction of the change of the radial variables on
the corresponding {\em physical} interval outside the source:
$\zeta \in (0, \infty) \leftrightarrow
z\in (1/(1-\varrho^2), \infty)$,
is a regular one.

It become obvious that the Hilbert solution \eqref{HilbertSol} and the solution \eqref{sol_NewGauge}
belong to mathematically and physically different gauge sectors of GR as a nonlinear gauge theory.

\section{Final form of the GR solution of the Einstein equations with single massive point source}

Expressions (\ref{sol_NewGauge}) and (\ref{rho_NewGauge})
still depend on the choice of the units for the new auxiliary variable $r$.

We must fix the scale $\tilde r$ in the form
\ben
\tilde r=\rho_G/\ln(1/\varrho^2)=G_NM/\ln\left({{M_0}\over{M}}\right)
\la{rscale}
\een
to ensure the validity of the standard asymptotic expansion:
$$g_{tt}\sim 1 - \rho_G/r +{\cal O}\left((\rho_G/r)^2\right)$$
when $r\to\infty$, i.e., the Eddington condition \cite{Eddington}.

This way the GR results for a single massive point particle,
as described in the new auxiliary Euclidian 3D space ${\cal E}^{(3)}$,
can be correctly compared with the results of the same problem
in the Newton theory of gravity in the flat 3D space ${\cal E}^{(3)}$.

Now in the basic regular gauge $\bar\varphi(r)=0$ the solutions for a massive point particle read
\ben
\varphi_1(r)=-{{G_N M}\over {r+G_N M}/\ln({M_0\over M})}=\varphi_{\!{}_G}(r),
\la{Gpot}
\een
\ben
\varphi_2(r)= \ln\left({\frac {2} {e^{-\varphi_G}-e^{\varphi_G}}}\right).
\la{phi2final}
\een

Using the modified (Newton-like) gravitational potential $\varphi_{\!{}_G}(r)$ \eqref{Gpot}, we can write down
the final form of the 4D interval defined by the new regular solutions:
\ben
ds^2=e^{2\varphi_{\!{}_G}(r)}
\left(dt^2-{{dr^2}\over{N_{\!{}_G}(r)^4}}\right)
-\rho(r)^2 d\Omega^2.\hskip .2truecm
\la{New_metric}
\een
where the coefficient
\ben
N_{\!{}_G}(r)=\left(2\varphi_{\!{}_G}\right)^{-1}
\left(e^{2\varphi_{\!{}_G}}-1\right)>0,
\la{NG}
\een
and the optical luminosity distance
\ben
\rho(r)=\rho_G/\left(1-e^{2\varphi_{\!{}_G}}\right)
={{r+G_NM/\ln({M_0\over M})}\over {N_{\!{}_G}(r)}}.
\la{rhoG}
\een

These basic formulas describe in a more usual way our regular solutions
of the Einstein equations for  $r\in [0,\infty)$ and show that:

1. In GR we have two types of regular solutions of the massive point particle problem:

A) Normal solutions for which $\varrho \in (0,1)$ and binding energy is positive $E_{binding}=M_0-M > 0$.

For them the modified newtonian gravitational potential \eqref{Gpot} is a regular bounded
function in the whole interval $r\in [0,\infty)$ without singularity at the origin $r=0$.
Its singular point $r_{singular}=\ln \varrho <0 $ is in the nonphysical domain $r < 0$.

In the language of Newton's gravity, the positivity of binding energy
reflects the fact that the effective gravitational force
is an attractive one and has a negative potential energy.

B) Abnormal solutions for which $\varrho \in (1,\infty)$ and binding energy is negative
$E_{binding}=M_0-M \leq 0$.

For them the modified newtonian gravitational potential \eqref{Gpot} is a singular function in the physical domain
$r \in [0,\infty)$ and its singular point is $r_{singular}=\ln \varrho \geq 0 $.
For the value $\rho=1$ we have $r_{singular}=0$, as in the case of Newton's gravity.

Despite the fact that the effective gravitational force for abnormal solutions is also an attractive one,
the positivity of potential energy in the interval $r \in [0,r_{sing})$ (in Newtonian language),
as well as the positivity of binding energy, together make transparent
the nonphysical character of the abnormal solutions.

Therefore, we do not consider them in more detail in the present paper.

2. For normal regular solutions $\varphi_G(0)= \ln\varrho <0$ is finite and
$$ds^2|_{r=0}=\varrho^2 dt^2- {\frac{\varrho^2\left(\ln\varrho\right)^4}{\left(1-\varrho^2\right)^4}}dr^2
-{\frac {\rho_G^2}{\left(1-\varrho^2\right)^4}} d\Omega^2.$$

3. In the limit $\varrho\to 1$ for any value of $r \in [0,\infty)$
we obtain $\varphi_{\!{}_G}\left(r;M,M_0\right) \to 0$ and at all 3D space points
$$g_{tt}(r)\to 1,\,\,\, g_{rr}(r)\to -1,\,\,\, \rho(r)\to \infty.$$

Due to the last result, the 4D geometry does not have a  meaningful limit when $\varrho\to 1$, i.e.,
massive point solutions without mass defect are not admissible in GR.

4. In the limit $\varrho\to 0$ our normal solutions tend to the Kohler-Pugachev-Gun'ko-Menzel ones,
see Section IV.A.9. In this case, as in the Newton gravity
$\varphi_{\!{}_G}\left(0\right)=\ln\varrho \to -\infty$.

5. For both normal and abnormal solutions in the limit $r \to r_{sing}$ we obtain
$$\lim_{r \to r_{sing}} ds^2 = \rho_G^2 d\Omega^2.$$

This result shows that the singular 2D surface $r = r_{sing}$ in $M^{(3)}\{-g_{mn}\}$ is not an event horizon.

It is not hard to see that the new class of regular solutions does not contain the Hilbert one, see Section IV.A.2.
The Hilbert solution cannot be considered also as a limit of any kind of the new regular solutions.

Thus we see that the Schwarzschild BH solutions and our regular solutions belong to different gauge sectors of GR.

\section{Total energy of the system of a point source and its gravitational
field}

In the problem at hand we have an extreme example of an "island
universe``. In it a privileged reference system and a well defined
global time exist.

It is well known that under these conditions
the energy of the gravitational field can be defined unambiguously
\cite{books}.

Moreover, we can calculate the total energy of the
aggregate of a mechanical particle and its gravitational field.

Indeed, the canonical procedure produces the total Hamilton density
\ben
{\cal H}_{tot}=\Sigma_{a=1,2;\mu=t,r}\,\pi_a^\mu\,\varphi_{a,\mu}-{\cal L}_{tot}= \nonumber \\ ={1\over{2G_N}}\left(-\bar\rho^2{\varphi_1^\prime}^2
+\bar\rho^2{\varphi_2^\prime}^2-e^{2\varphi_2}\right)+M_0
e^{\varphi_1}\delta(r).
\la{Htot}
\een

Using the constraint (\ref{epsilon}) and
the first of the relations (\ref{In_constMM}), one immediately obtains
for the total energy of the GR universe with one point particle in
it:
\ben E_{tot}=\int_0^{{\infty}}{\cal H}_{tot} dr=M=\varrho
M_0\leq M_0\,.\la{E}\een

This result completely agrees with the strong equivalence
principle of GR. The gravitational energy, i.e., the total energy of the gravitational field,
created by a point particle, is the negative quantity:
$$E_{gr}=E_{tot}-E_0=M-M_0=-M_0(1-\varrho)<0.$$

The above consideration gives a clear physical explanation of the
gravitational mass defect for a point particle and justifies the physical meaning of our normal regular solutions.

\section{Invariant Characteristics of the Solutions with a Point Source}
\subsection{Local Singularities of Point Sources}
Using the invariant 1D Dirac function, one can write down the first
of Eqs. (\ref{FEq}) in the form of the {\em exact relativistic}
Poisson equation: \ben \Delta_r \varphi_1(r)=4\pi G_N
M_0\,\delta_g(r).\la{Poisson}\een This equation is a specific
realization of Fock's idea (see Fock in \cite{books}) using our
normal field coordinates (Section VII). It can also be rewritten
in transparent 3D form:
\ben
\Delta \varphi_1(r)=4\pi G_N
M_0\,\delta_g^{(3)}({\bf r}).
\la{3DPoisson}
\een

The use of the invariant  Dirac $\delta_g$-function has the advantage that
under diffeomorphisms of the space  ${\cal M}^{(3)}\{-g_{mn}({\bf r
})\}$ the singularities of the right-hand side of the relativistic
Poisson equation (\ref{Poisson}) remain unchanged. Hence, we have
the same singularity at the place of the source for the whole
class of physically equivalent rho-gauges.

Then one can distinguish the physically different solutions of the Einstein
equations (\ref{Einst}) with spherical symmetry by investigating
the asymptotic in the limit $r\to +0$ of the coefficient
$$\gamma(r)={\frac {1}  {4\pi\rho(r)^2\rho^\prime(r)\sqrt{-g_{rr}(r)}}}=
{\frac {\sqrt{1-\rho_G/\rho(r)}}  {4\pi\rho(r)^2{{d\rho}\over{dr}}}}$$
in front of the usual 1D Dirac $\delta(r)$-function in the representation
$\delta_g(r)=\gamma(r)\delta(r)$ of the invariant one.
\begin{figure}[ht!]
\vskip -.truecm
\hskip -0.truecm
\begin{minipage}{10cm}
\includegraphics[width=0.8\textwidth]{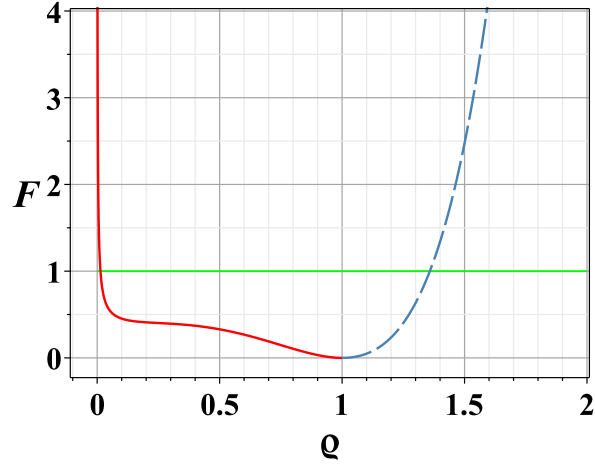}
\vskip .truecm
\caption{The graph of the universal function $F(\varrho)$. The solid red line ($\varrho \in (0,1)$) represents the physical (normal) branch.
The dashed blue line ($\varrho \in (1,\infty)$) represents the nonphysical (abnormal) branch.
The intersection of some given level-line (horizontal dashed-dot green)
with the graph indicates the two roots $\varrho_{physical}<1$ and $\varrho_{nonphysical}>1$.}
\label{fig1}
\end{minipage}
\end{figure}

For regular solutions (\ref{sol_f}), (\ref{sol_NewGauge}) the
limit $r\to +0$ of this coefficient is a constant
\ben
\gamma(0)={1\over {4\pi \rho_G^2}} \,{\frac { \left(1 -\varrho^2\right)^4} {\varrho\left(2\ln\varrho\right)^2}}=
{1\over {4\pi \rho_G^2}} \,F(\varrho),
\la{gamma}
\een
which describes the intensity of the invariant $\delta$-function and leads to the formula
$\delta_g(r)=\gamma(0)\delta(r)$. The graph of the universal function $F(\varrho)$ is shown in Fig.1.

Now we see that:

1) The condition $\varrho\in (0,1)$ ensures a correct sign of the
intensity $\gamma(0)$, i.e., the property
$\gamma(0)\in(0,\infty)$, and thus, the negativity of the total
energy $E_{gr}$ of the gravitational field of a point particle,
according the previous Section XII.

2) For $\varrho=0$ we have $\gamma(0)=\infty$, and for
$\varrho\!=\!1$:  $\gamma(0)\!=\!0$. Hence,  in these two cases we have non-physical
values of the intensity $\gamma(0)$.

Once more, our conclusion is that the {\em physical} interval of values of the
ratio $\varrho$ for regular solutions is the open interval  $(0,1)$. This is consistent
with the mathematical analysis of this problem, given in the
previous sections.

Consider the same situation for other known solutions of the Schwarzschild problem,
listed in Section IV.A.

i) For the original the Schwarzschild solution one obtains
$$\gamma_{{}_{Schwarzschild}}(0)=\infty.$$
The corresponding Eq.\eqref{gamma} gives two nonphysical solutions: $\varrho_1=0$ and $\varrho_2=\infty$, see Fig. 1.
Hence, this solution for a massive point particle is not a physical one but may be considered as a limiting case of the physical solution.

ii) As a result of the alteration of the physical meaning of the
variable $\rho=r$ inside the sphere of the radius $\rho_{{}_H}$, in
the Hilbert gauge the coefficient $\gamma(r)$ tends to {\em
imaginary} infinity for $r\to 0$:
$$\gamma_{{}_{Hilbert}}(r)\sim {{i\rho_G}\over {4\pi r^{5/3}}} \to i \infty. $$

This is in sharp contrast to the {\em real} asymptotic of all other solutions at hands and shows once more that the Schwarzschild BH belong to a different gauge class
of the Schwarzschild problem.

iii) The Droste solution is a regular one:
$$\gamma_{{}_{Droste}}(0)={1\over {4\pi \rho_G^2}}$$
From Eq.\eqref{gamma} one obtains the physical value of the gravitational mass defect
$$\varrho_{{}_{Droste}}\approx 0.0134612785,$$
which belongs to the normal physical interval $\varrho\in (0,1)$
and shows that the Droste solution is a regular one and
can be transformed to (\ref{sol_f}), or (\ref{sol_NewGauge}) by a proper
regular rho-gauge transformation.

There exists another root of the same equation $\varrho_{{}_{Droste}} \approx 0.0134612785 $,
which belongs to the abnormal branch and does not correspond to a
physical solution.

iv) For the Einstein-Rosen solution, as a solution for a massive point particle, one obtains the nonphysical value
$$\gamma_{{}_{Einstein-Rosen}}(0)=0$$
and nonphysical value of the mass defect ratio $\varrho_{{}_{Einstein-Rosen}}=0$.

v) The case of the Weyl solution is more complicated.
This solution resembles a nonphysical regular solution with
$\gamma_{{}_{Weil}}(0)=0$
and nonphysical value of the mass defect ratio $\varrho_{{}_{Weil}}=0$.

Actually, the exact invariant $\delta$-function for this solution can be written in
the form (see the Appendix B):
\ben
\delta_{g,{}_W}(r)={{16}\over{\pi \rho_G^2}} \left(
\Big({{r}\over{\rho_G}}\Big)^4\delta(r)+
\left({{\rho_G}\over{r}}\right)^4\delta\Big({{\rho_G^2}\over
r}\Big) \right).
\la{Wdelta}
\een

This distribution acts as a zero-functional on the standard class of smooth test functions
with compact support, which are finite at the points $r=0,\infty$.

Formula (\ref{Wdelta}) shows that:

a) The Weyl solution describes a problem with  spherical symmetry
in the presence of {\em two} point sources: one with
$\varrho_{{}_W}^0=1$ at the point $r=0$ and the other one with
$\varrho_{{}_W}^\infty=1$ at the point $r=\infty$.

Thus, this solution is the first {\em exact} analytical
two-particle-like solution of the Einstein equations (\ref{Einst}).

Unfortunately, as a two-point solution, the Weyl solution is
physically trivial: it does not describe the dynamics of two point
particles at a finite distance, but rather gives only a static
state of two particles at infinite distance.

b) To make these statements correct, one has to compactify the
comformally flat 3D space ${\cal M}^{(3)}\{-g_{mn}({\bf r })\}$
joining it to the infinite point $r=\infty$. Then, both the Weyl
solution {\em and} its source will be invariant with respect to
the inversion $r \rightarrow \rho_G^2/r$.

vi) The singularity of the coefficient $\gamma(r)$ in front
of the usual 1D Dirac $\delta$-function in the coherent gauge
solution is stronger than any other of listed solutions:
$$\gamma_{{}_{coherent}}(0)=\infty.$$
The corresponding Eq.\eqref{gamma} gives two nonphysical solutions: $\varrho_1=0$ and $\varrho_2=\infty$, see Fig.1.
Those solutions can be considered as a limiting case of the regular physical solutions (\ref{sol_f}), or (\ref{sol_NewGauge}).

\subsection{Invariants of the Space-Times of Single Massive Point Particle}
It is well known that the scalar invariants of the Riemann tensor
allow a manifestly coordinate independent description of the
geometry of space-time manifold.

The problem of a single point
particle can be considered as an extreme case of a perfect fluid,
which consists of only one particle.

According to the article
\cite{CMcL}, the maximal number of independent {\em real}
invariants for such a fluid is 9. These are the scalar curvature $R$
and the standard invariants $r_1, w_1, w_2, m_3, m_5$. (See
\cite{CMcL} for their definitions.)

The invariants $R$, $r_1$ and
$m_3$ are real numbers and the invariants $w_1$,$w_2$ and $m_5$
are, in general, complex numbers.  The form of these invariants
for spherically symmetric space-times in normal field
variables is presented in Appendix C. Using these formulas we
obtain for the regular solutions (\ref{sol_f}) in the basic gauge
the following simple invariants:
\begin{subequations} \label{C:1,2,3,4}
\ben
I_1&=&{1\over{8\rho_G^2}}{{(1-\varrho^2)^4}\over{\varrho^2}}
\delta\left({r\over{\rho_G}}\right)=\nonumber \\
&=&{{\pi(1-\varrho^2)^3}\over{2\varrho}}
\delta_g\left({r\over{\rho_G}}\right)=-{1\over 2}R(r),  \label{C:1} \\
I_2&=&0, \label{C:2} \\
I_3&=&{{\Theta(r/\rho_G)-1}\over{8\rho_G^2\varrho^2}}\left(1-\varrho^2
e^{4r/\rho_G}\right)^4=\nonumber \\
&=&{{\rho_G^2}\over{8\varrho^2}}{{\Theta(r/\rho_G)-1}\over{\rho(r)^4}}, \label{C:3} \\
I_4&=&{{\Theta(r/\rho_G)}\over{4\rho_G^2}}\left(1-\varrho^2
e^{4r/\rho_G}\right)^3\!=\nonumber \\
&=&{{\rho_G}\over{4}}{{\Theta(r/\rho_G)}\over{\rho(r)^3}}.  \label{C:4}
\een
\end{subequations}

During the calculation of the expressions $I_{3,4}$ we have used
once more our assumption $\Theta(0)=1$.

As can be easily seen:

1) The invariants $I_{1,...,4}$ of the Riemann tensor are well-defined distributions and:

i) This is in accordance with the general
expectations described in the articles \cite{TaubRaju}, where one
can find a correct mathematical treatment of distributional valued
curvature tensors in GR.

As we saw, the manifold ${\cal M}^{(1,3)}\{g_{\mu\nu}(x)\}$ for
our regular solutions has a geometrical singularity at the point,
where the physical massive point particle is placed.

ii) The metric for the regular solutions
is globally continuous, but its first
derivative with respect to the radial variable $r$ has a finite jump
at the point source.

This jump is needed for a correct
mathematical description of the delta-function distribution of
matter of the massive point source in the right-hand side of
the Einstein equations (\ref{Einst}).

iii) Three of the invariants \eqref{C:1,2,3,4} are independent for the values $r\in (-\infty,\infty)$ on the
real axes and this is the maximal number of
the independent invariants in the problem of a single point source
of gravity.

On the real physical interval $r\in [0,\infty)$ one
has $I_3=0$ and we remain with only two independent invariants.

For $r\in (0,\infty)$ the only independent invariant is $I_4$, as
is well known from the case of the Hilbert solution.

2) The scalar invariants $I_{1,...,4}$ have the same form as in
Eq. \eqref{C:1,2,3,4} for the regular solutions in the
representation (\ref{sol_NewGauge}) obtained via the fractional linear mapping (\ref{rrgt}).

3) All other geometrical invariants $r_1(r)$, $w_{1,2}(r)$,
$m_{3,5}(r)$ include degrees of the Dirac $\delta$-function and are
not well defined distributions.

Therefore, the choice of the simple
invariants $I_{1,...,4}$ is essential and allows us to use
solutions of the Einstein equations, which are distributions.

To the best of author's knowledge, the requirement to make
the correct use of the theory of distributions  possible is a novel criteron
for the choice of invariants of the curvature tensor and seems
not been used until now.

It is curious to know whether it is
possible to apply this new criterion in more general cases, which
differ from the case of static spherically symmetric space-times.

\subsection{Existence of Event Horizon for Static Spherically Symmetric Space-Times}

According to the well-known theorems by Hawking, Penrose, Israel
and many other investigators, the only solution with a regular
event horizon not only in GR but in the theories with scalar
field(s) and in more general theories of gravity is the Hilbert
one \cite{FN}.

As we have seen, this vacuum solution has a pure
geometrical nature and does not describe a gravitational field of
a point particle with proper mass $M_0\neq 0$.

The strong mathematical results like the well-known no hear theorems, etc,
for the case of metrics with an event horizon are
based on the assumption that such horizon indeed exists.

These mathematical results do not
contradict the ones obtained in the present article, because
the other solutions, which we have considered together with the
Hilbert one, do not have an event horizon at all.

Indeed, for the point  $\rho_{{}_H}$ at which
$g_{rr}(\rho_{{}_H})=\pm\infty$ one obtains the equation
$$\rho(\rho_{{}_H})=\rho_G.$$
The last equation does not have any solution $\rho_{{}_H}\in {\cal
C}^{(1)}$ for the regular solutions (\ref{sol_f}),
(\ref{sol_NewGauge}). The absence of a horizon in the physical
domain $r\in [0,\infty)$ is obvious in the representation (\ref{New_metric})
of the regular solutions.

For the other classical solutions one obtains from the above equation

Schwarzschild solution:\hskip 3.3truecm  $\rho_{{}_H}=0$;

Hilbert solution: \hskip 4.2truecm $\rho_{{}_H}=\rho_G$;

Droste solution: \hskip 4.26truecm $\rho_{{}_H}=0$;

Weyl solution: \hskip 4.54truecm $\rho_{{}_H}=\rho_G$;

Einstein-Rosen solution: \hskip 3.truecm $\rho_{{}_H}=0$;

Isotropic (t-x) solution: \hskip 3.15truecm $\rho_{{}_H}=-\infty$;

Coherent gauge solution: \hskip 2.95truecm  \hskip 0.truecm $\rho_{{}_H}=0$.

As one can see, only for the Hilbert and Weil
solutions the metric component $g_{rr}(r)$ has a simple pole at
the finite point $\rho_{{}_H}=\rho_g$, i.e., the solution has an event horizon at the finite point in the physical domain .

In the cases when $\rho_{{}_H}=0$, we have a physical massive point particle at this point, if the mass ratio $\varrho\neq 0,1$.

%%%%%%%%%%%%%%%%%%%%%%%%%%%%%%%%%%%%%%%%%%%%%%%%%%%%%%%%%%%%%%%%%%%%

\section{Unusual 3D Geometry of Massive Point Source in GR}

It is curious that for a metric given by a regular solution the
3D-volume of a ball with a small radius $r_b<\rho_G$ centered at
the source is
$$Vol(r_b)={4\over 3} \pi \rho_G^3\,
{{12\varrho}\over{(1-\varrho^2)^4}} {{r_b}\over{\rho_G}}+ {\cal
O}_2({{r_b}\over{\rho_G}}).$$

It goes to zero linearly with
respect to $r_b\to 0$, in contrast to the Euclidean case, where $Vol_E(r_b)\sim r_b^3$.

This happens, because for the regular
solutions (\ref{sol_f}), (\ref{sol_NewGauge}) in the limit $r \to 0$ we obtain
$$\sqrt{|{}^3g|}=\rho(r)^2\sin\theta \to \rho_{min}^2\sin\theta \neq 0.$$

Nevertheless,
$$\lim_{r_b\to 0}\,Vol(r_b)=0$$
and this legitimates the use of the term "a point
source of gravity" in the problem at hand:
the source can be surrounded by a sphere with an
arbitrary small volume $Vol(r_b)$ in it and with an arbitrary
small radius $r_b$.

In contrast, when $r_b\to 0$ the area of the ball surface has a
finite limit
$$\lim_{r_b\to 0}\,A(r_b)={{4\pi \rho_G^2}\over {(1-\varrho^2)^2}}> 4\pi \rho_G^2,$$
and the length of a big circle on this surface tends
to a finite number
$$\lim_{r_b\to 0}\,l(r_b)={{2\pi \rho_G}\over{1-\varrho^2}}>2\pi \rho_G.$$

Such unusual geometry, created by the massive point sources in GR,
may have interesting physical consequences.

For example, spacetimes
defined by the regular solutions (\ref{sol_f}) and
(\ref{sol_NewGauge}) have a unique property: when one approaches
the point source, its luminosity remains finite, see also the formula \eqref{A}.

This leads to an extremely important  modification of the Gauss theorem in the corresponding
3D spaces.

After all, this modification may solve the well-known
problem of the {\em classical} divergences in the field theory,
because, as we saw, GR offers a natural cut-off parameter
$\tilde r=\rho_G/\ln(1/\varrho^2)$ for the
fields created by massive point particles.

If this will be confirmed by more detailed calculations, the price one must pay
for the possibility to overcome the old classical divergences problem will be to accept the fact
that the massive point objects in GR can have a nonzero surface, having at the same time
a zero volume and a zero size: $r=0$ and $l=0$ .

On the other hand, as we have seen,  the inclusion of point
particles in GR is impossible without such unusual geometry.

The Hilbert solution no longer offers an attractive alternative
for description of a source of gravity as a physical massive {\em
point} placed  at the coordinate point $\rho=0$ and  with {\em
usual} properties of a geometrical point.

Hence, in GR we have no
possibility to introduce the notion of a physical point particle
with nonzero proper mass without a proper change of the standard
definition for a point particle as a mathematical object, which has
simultaneously zero size, zero surface and zero volume.

In GR we are forced to remove at least one of these three usual requirements as
a consequence of the concentration of the proper mass $M_0>0$ at only one
3D space geometrical point.

From a physical point of view the novel properties of a massive physical point
particle, at which we arrived in this article, is obviously
preferable.

Moreover, it seems natural to have an essential
difference between the geometry of  the ``empty'' space-time
points and that of the matter points in GR since in the last case, keeping the finite proper mass $M_0>0$ into one geometrical point, we will obviously have an infinite mass density at this geometrical point.

According to the basic idea of GR, such an infinite mass density must produce an essentially strong deformation of the spacetime and, as we have shown here in detail, it creates a very unusual spacetime geometry.

The squared 4D infinitesimal distance at this point becomes
$$ds^2|_{r=0}=\varrho^2 dt^2- {\frac{\varrho^2\left(\ln\varrho\right)^4}{\left(1-\varrho^2\right)^4}}dr^2
-{\frac {\rho_G^2}{\left(1-\varrho^2\right)^2}} d\Omega^2.$$

 This formula shows explicitly also the existence of a conic singularity with an angle excess around the massive point particle source of the gravitational field.

Alike geometry for such essentially different objects like an empty mathematical point and a massive physical point is possible only in the classical physics, where the spacetime geometry does not depend
on the matter. Only under the last assumption there is no {\em geometrical} difference between pure mathematical space points and matter points.

The intriguing novel situation in GR, described in detail in the present
article, deserves further careful analysis.

The appearance of the above nonstandard geometry in the point mass
problem in GR was discussed without much details for the first time by Marcel Brillouin \cite{Brillouin}.

Our detailed consideration of the regular solutions (\ref{sol_f}),
(\ref{sol_NewGauge}) confirms his preliminary guess on the character
of the singularity of the gravitational field of massive point
particles in GR.

%%%%%%%%%%%%%%%%%%%%%%%%%%%%%%%%%%%%%%%%%%%%%%%%%%%%%%%%%%%%%%%%%%%

\section{Concluding Remarks}

The most important result of the present article is the explicit
demonstration of the fact that there exist infinitely many different static
solutions of the Einstein equations (\ref{Einst}) with spherical symmetry,
a mass point singularity placed at the center of symmetry,
and vacuum outside this singularity,  and with the same Keplerian mass $M$.

These solutions fall into different gauge classes, which
describe physically and geometrically different space-times.

Several of them were discovered at the early stages
of the development of GR. Up to now they are often considered as
equivalent representations of some "unique" solution, which depends on
only one parameter -- the Keplerian mass $M$.

As shown in the present paper, this is not the case.

As a consequence of the Birkhoff theorem, when expressed in the same variables,
the solutions with the same mass $M$ indeed coincide in their common
domain of coexistence -- outside the singularities.

In contrast, they may have different behavior at the corresponding
singular points.

A correct description of a massive point source of gravity is impossible,
if using most of these classical solutions.
The only exceptions are the Droste solution and the coherent gauge solution introduced in the present paper.

Using novel normal coordinates for the gravitational field of a single
point particle with the proper mass $M_0>0$, we were able to
describe correctly the massive point source of gravity in GR. The
singular gauge transformations yield the possibility to overcome
the restriction to have a zero proper mass $M_0=0$ for neutral point
particles in GR \cite{ADM}. It turns out that this problem has a
two-parametric family of regular solutions.

One of the parameters -- the proper mass $M_0$ of the point source,
can be obtained in the form of surface integral, integrating both
sides of Eq. (\ref{3DPoisson}) on the whole 3D space
${\cal M}^{(3)}\{-g_{mn}\}$:

\ben M_0&=&{1\over{4 \pi
G_{{}_N}}}\int_{{\cal M}^3}d^3{\bf r} \sqrt{|{}^3g|}\,\Delta
\left(\ln\sqrt{g_{tt}}\right)=\nonumber \\
&=&{1\over{4 \pi
G_{{}_N}}}\oint_{\partial{\cal M}^3}d^2\sigma_i
\sqrt{|{}^3g|}\,\,g^{ij}\,\partial_j \left(\ln\sqrt{g_{tt}}\right).\nonumber\een

The second parameter is the  Keplerian mass $M$ described,
as shown in Section XII, by the total energy:
$$M=\int_{{\cal M}^3}d r\,{\cal H}_{tot}=
\int_{{\cal M}^3}d^3{\bf r}\sqrt{|{}^4g|}\,M_0\, \delta_g^{(3)}({\bf r}).$$

These two parameters define the {\em gauge class} of the given regular solution.

Obviously, the parameters $M$ and $M_0$ are invariant
under regular static gauge transformations, i.e., under diffeomorphisms
of the 3D space ${\cal M}^{(3)}\{-g_{mn}\}$. It is an analytical
manifold with a strong singularity at the place of the massive point
particle.

For every regular solution of the Einstein equations \eqref{Einst} both parameters are finite and positive and
satisfy the additional physical requirement $0<M<M_0$, which ensures the negative energy of the gravitational field.

It is convenient to use a more physical set of continuous
parameters for fixing the regular solutions, namely:
the Keplerian mass $M\in (0,\infty)$ and the  gravitational
mass ratio $\varrho={{M}\over{M_0}}\in(0,1)$.

A classical solution, which is regular, is the Droste
one. For this solution the gravitational mass ratio is
$\varrho_{{}_{Droste}}\approx 0.0134612785$.

For the regular solutions, the physical values of the optical
luminosity distance $\rho$ are in the semi-constraint interval
$\rho\in [{{\rho_G}\over{1-\varrho^2}}, \infty)$.

For small values of the mass ratio $\varrho \to +0$ our regular massive point solutions of the Einstein equations violate the Buchdahl theorem \cite{Buchdahl}.
According to this theorem, the finite dimensional massive spherically symmetric solutions with the Keplerian mass $M$, like stars, cannot have luminosity radius $\rho_* < \rho_{{}_{Buchdahl}}={\frac 9 8} \rho_G$.

As correctly pointed out in \cite{Cardoso2019}, mathematical theorems are valid only if the mathematical assumptions in their formulation are fulfilled.

In the paper \cite{Cardoso2019},  six such assumptions, on which the Buchdahl theorem is based, are listed. Their violation opens the door to the development of the theory of ECOs.

In this list, the most important {\em geometrical} assumption, which forbids the values $\rho_* \in (\rho_G,\rho_{{}_{Buchdahl}})$ is missing.

This is the assumption that the values of the luminosity distance inside the spherically symmetric matter objects belong to the interval $\rho \in (0,\rho_* )$. This assumption seems natural if one wrongly supposes that in the curved spacetime the luminosity distance can be used instead of the real geometrical distance. Thus, the unusual geometries described in Section XIV are missing.
The results of the present paper are based on the exclusion of this assumption of the Buchdahl theorem.

Outside the source, i.e., for $\rho>{{\rho_G}\over{1-\varrho^2}}$,
the Birkhoff theorem is strictly respected for all regular
solutions.

The metric defined by a regular solution is {\em globally}
continuous in the whole domain $r \in {-\infty,\infty}$, but its first derivatives have a jump at the position of the matter point source. The second derivative of the metric is a distribution described by the Dirac function.

Our explicit results justify the Raju conclusion
\cite{TaubRaju} that GR will be consistent with the existence of point
particles if one assumes the metric to be at most of class $C^0$
and show that the Hawking-Penrose singularity theory,
based on the assumption that the components of $g_{\mu\nu}$ are
{\em at least} $C^1$ functions, must be reconsidered.

For the class of regular solutions written in the form (\ref{sol_f}),
(\ref{sol_NewGauge}), or (\ref{New_metric}),
the non-physical interval of the optical
luminosity distance $\rho\in[0,{{\rho_G}\over{1-\varrho^2}}]$,
which includes the luminosity radius $\rho_{{}_H}=\rho_G$, must be
considered as an "optical illusion".

All pure mathematical objects, which belong to this interval, are in the imaginary
domain "behind" the real physical source of the gravitational
field and have to be considered as a specific kind of optical "mirage".

This is in agreement with Dirac's conclusion about the
non-physical character of the domain $\rho \in [0, \rho_G] $  \cite{Dirac} for point particles and extends this conclusion to
the whole non-physical interval of the optical luminosity distance
$\rho\in[0,{{\rho_G}\over{1-\varrho^2}}] \supset [0, \rho_G] $ for a given
{\em regular} solution.

We are forced to cut the Hilbert form of the vacuum Schwarzschild
solution at the value $\rho_{min}={{\rho_G}\over{1-\varrho^2}}>\rho_G$
because of the presence of the matter point mass. Its presence
ultimately requires a definite jump of the first derivatives of
the metric. The jump is needed to make the Einstein tensor
coherent with the energy-momentum stress tensor of the point
particle. This cutting can be considered as a further development and physical clarification
of Dirac's idea \cite{Dirac}:

{\em ``Each particle must have a finite
size no smaller than the Schwarzschild radius. I tried for some
time to work with a particle with radius equal to the
Schwarzschild radius, but I found great difficulties, because the
field at the Schwarzschild radius is so strongly singular, and it
seems that a more profitable line of investigation is to take a
particle bigger than the Schwarzschild radius and to try to
construct a theory for such particle interacting with the
gravitational field.''}

For correct understanding of these statements one has to take
into account that the above Dirac idea is expressed just in terms
of the luminosity radius $\rho$.

Excluding the auxiliary variable $r$ from formulas \eqref{nc}, \eqref{sol_f} and \eqref{basic_rho_f},
for massive point particle in GR we obtain solutions in a form
$g_{tt}(\rho)=1-{\frac {\rho_G} {\rho} }$ and $ g_{\rho\rho}(\rho)=\left(1-{\frac {\rho_G} {\rho} }\right)^{-1}$,
which is valid {\em only} for $\rho > {\frac {\rho_G}{1-\varrho^2}}=\rho_{min}$.
This Hilbert-gauge-like form is derived on the basis of our detailed considerations and
deciphers the precise meaning of the Dirac statement.

An alternative development, which is much closer to the Dirac approach
to the above problem, can be found in the recent articles \cite{MM},
as well as in the latest
developments \cite{Cardoso16,Mark2017,Cardoso2017,Conklin18,Wang2018,Raposo18,Cardoso2019}, where
the general framework of ECOs, which are not BH, is developed in relation with possible echoes in gravitational waves from binary merger.
Our approach sheds essentially new light on these new problems.

The geometry around physical massive point particles is
essentially different from the geometry around the "empty"
geometrical points.

This unusual geometry, discussed for the  first time by Brillouin, may offer a new way for overcoming the {\em
classical} fields divergence problems. The new possibility differs
from the one used in \cite{ADM} for charged point particles. It
is based on the existence of the natural cut-off parameter
$\tilde r=G_{\!{}_N}M\left(c^2\ln\left({{M_0}\over M}\right)\right)^{-1}$
and on a new interpretation of the  relation between the
singular mass distribution of a point particle and the geometry of
the space-time around this particle.
This relation is described here for the first time in full detail (See also \cite{Fiziev2003}.).

According to the remarkable comment by Poincar\'e \cite{Poincare},
real problems can never be considered as solved or unsolved
but rather they are always {\em more or less} solved.

The strong {\em physical} singularity at the "event horizon" of
the Hilbert form of solution of the Schwarzschild problem in the presence of a massive point source,
stressed by Dirac, as well as further physical consequences, which one can derive from the new
regular solutions (\ref{sol_NewGauge}) of this more than one century old problem,
will be discussed in separate articles, see, for example \cite{Fiziev2007}.

The corresponding development for a charged massive point source can be found in \cite{FizievDimitrov2004}.

Here we will add some more remarks:

Our consideration shows that the observed by the astronomers
extremely compact dark objects ECOs, called by them currently black {\em holes} (BHs)
without any direct evidences for the existence of {\em real} event horizons \cite{AKL},
cannot be described theoretically by solutions
(\ref{HilbertSol}) in the Hilbert gauge, if we assume that these objects are made of matter of some nonzero proper mass $M_0>0$.

At present, the only real fact is the existence of  massive
invisible ECOs with the Keplerian mass $M\gtrapprox 5 M_\odot$, which is too large with respect to
the conventional understanding of the star physics.

Concerning these unusual ECOs, most probably one actually has to
solve a much more general problem.

Since a significant amount of invisible dark matter, which manifests
itself only due to its gravitational field, is observed in the
Nature at very different scales: in the clusters of galaxies, in
the halos of the galaxies, at the center of our galaxy, and  as
compact dark components in some binary star systems, one is
tempted to look for some universal explanation of all these
phenomena. Obviously, such universal explanation cannot be based
on the Schwarzschild solution in the Hilbert gauge and one has to look for
some other theoretical approach. A similar idea was pointed out
independently in the recent articles \cite{RN}.

At the end, we wish to mention some additional open problems, both
mathematical and physical, connected with the gravitational field of point particles.

Our consideration was essentially restricted to the real domain
of variables. The only exception was the description of the
singular change (\ref{change}) in the radial variable.

It is well known that the natural domain
for studying the solutions of holomorphic differential equations
is the complex one. Complete knowledge of
solutions of these equations is impossible without a description of
all singularities of the solutions in the complex domain.

Therefore, looking for a complete analysis of the solutions of
some differential equation of $n$-th order
$f\left(w^{[n]}(z),w^{[n-1]}(z),...,w^{[1]}(z),w(z),z\right)=0$
for a function $w(z), z\in {\cal C}^{(1)}$, where $f(...)$ is a
holomorphic function of the corresponding complex variables, one
has to consider these solutions as holomorphic functions of the
complex variable $z$. Now the study of the singular points of the solutions in the {\em whole}
complex domain becomes the most important issue.

For the Einstein equations (\ref{Einst}) a full {\em four-dimensional}
complex analysis is impossible at present, because the corresponding
mathematical methods are not developed enough.

For the problem of single massive point source in GR, in its 1D formulation used in this article,
one can apply the classical complex analysis.
Moreover, using the well-known complex
representation of distributions (see Bremermann in \cite{Gelfand}),
it is not difficult to generalize the classical results for
the case when the differential equation has on the right-hand side
a distribution that depends on the variable $z$.
Such a term may lead to a discontinuity of the solution $w(z)$
or its derivatives.
The corresponding complex analysis of the solution remains
an important open mathematical problem, together with the problem
of finding and classifying {\em all} static solutions of the Einstein
equations with one matter point singularity and spherical symmetry.

The analysis of QNMs of waves of different spin on the Schwarzschild or Kerr BH background in complex domain of the radial variable $\rho$ was the basic idea for successful high-precision computations in \cite{Fiziev2006,Fiziev2009,Fiziev2011,Fiziev2010}.

To find criteria for
an experimental and/or observational probe of different solutions
of the Einstein equations with spherical symmetry becomes the most important physical problem.
Since there exist
{\em physical and geometrical} differences between the solutions
of this type, one can find a real way of distinguishing them
experimentally.

Especially, a problem of the present day is to
answer the question of which of the spherically symmetric
solutions gives the right description of the observed ECOs if one
hopes that the ECOs can be considered in a good approximation as
spherically symmetric objects with some Keplerian mass $M>0$ and some
mass ratio $\varrho\in (0,1)$.

In this case, according to the Birkhoff theorem,
we must use the new solutions (\ref{New_metric}) for describing
their gravitational field in the outer vacuum domain, outside the ECOs.

For the same reason, the solutions (\ref{New_metric}) must be used also
for describing the vacuum gravitational field around
spherically symmetric bodies of finite size $R$,
made of usual matter.

Due to the specific character
of the correction of the Newton law of gravitation
defined by the term $G_{\!{}_N}Mc^{-2}/\ln\left({{M_0}\over M}\right)$ in
the potential $\varphi_{\!{}_G}(r;M,M_0)$ (\ref{Gpot}),
the difference between our solutions  (\ref{New_metric}) and
the Hilbert one is in higher order relativistic terms, when
$\ln\left({{M_0}\over{M}}\right)\sim 1$.

This difference will not influence the rho-gauge invariant local
effects like gravitational red-shift, perihelion shift, deviation of light rays,
time-dilation of electromagnetic pulses, and so on, but it may be
essential for the quantities, which depend directly on the precise form of
the potential $\varphi_{\!{}_G}(r;M,M_0)$ (\ref{Gpot}) and especially on the minimal admissible value $\rho_{min}={\frac {\rho_G}{1-\varrho^2}}\geq \rho_G$.

For all bodies in the solar system the term
$G_{\!{}_N}Mc^{-2}/\ln\left({{M_0}\over M}\right)$ is very small in
comparison with their physical radius $R$:
$\left(\rho_{\!{}_G}/2R\right)_{Earth}\sim 10^{-10}$,
$\left(\rho_{\!{}_G}/2R\right)_{\odot }\sim 10^{-6}$, etc.

Nevertheless, the small differences between the Newton potential
and the modified one may be observed in the near future
in the high precision measurements  under the programs like
APOLLO, LATOR, etc. (see for example \cite{LATOR}) if we will
be able to find the proper quantities, which depend on the precise form of
the potential $\varphi_{\!{}_G}(r;M,M_0)$ (\ref{Gpot}).

A better possibility for observation of deviations from Newton's
gravitational potential and in four-interval may be offered by neutron stars, since
for them $\left(\rho_{\!{}_G}/2R\right)_{n*}\sim 0.17-0.35$.

The results of the present article may have an important impact on
the problem of gravitational collapse in GR and the experimental study of GW,
which is already accessible for us \cite {LOGOOPEN,Barack2018}.

Up to now most of the investigations known to the author presuppose to have BH as a
final state of the collapsing object, i.e. the Hilbert
solution, or its rotating Kerr generalization with a more general kind of the event
horizon, in the spirit of the Roger Penrose cosmic censorship hypotheses
and the Kip Thorn hoop conjecture.

These two conjectures have not yet been proven, after more than 50 years of hard efforts in this direction.
They both disregard the unusual spacetime geometry created by a massive point with a mass defect,
see Section XIV.

Both conjectures presuppose a jump between the solutions of the Einstein equations, which belong
to different gauge classes.
Such a topological jump cannot be obtained in a continuous way, for example,
during the exhaustion of a star's nuclear fuel, which is a continuous process from  a macroscopic point of view.

In the last decade, more attention was paid also to
the final states with necked singularities. The
existence of a two-parametric class of regular static solutions
without an event horizon opens a new perspective for these
investigations.

From a conceptual point of view,
one can stress that the existence of different gauge classes of solutions in GR,
which is certainly a nonlinear gauge theory,
is still not enough recognised and carefully used in different kinds of its applications.

Most probably, the proper understanding of the origin
of the mass defect of a point particle, introduced in the present
article, will be reached by a correct description of the
gravitational collapse.

Without any doubts, the inclusion of the new static spherically
symmetric regular solutions of the Einstein equations in the corresponding
investigations will open new perspectives for further
developments in GR, as well as in its modern generalizations.

A proper generalization of our approach to rotating massive-point-particle sources in GR is urgently needed.

Application of the new regular solutions for studying dynamics of several point particles in GR may also be of critical importance for present-day experimental applications of GR, especially in theoretical studies of GW, radiated by such systems of particles \cite{Jafari2019}.

\begin{acknowledgments}

I'm deeply indebted to prof. Ivan Todorov and to the participants
of his seminar for illuminating discussions, comments and
suggestions, to prof. Martin Reuter, Dr. Salvatore Antoci, Dr.~James Brian Pitts,
and to Dr. Lluis Bel, Dr. Bill Davis,  Dr. Stan Robertson
for drawing my attention to important references.

My special thanks are to Dr. Salvatore Antoci for sending me copies of
the articles  \cite{Dirac}.

I wish to thank also Dr. Vasil Tsanov for useful discussion of multi-dimensional complex analysis,
prof. G.~Schaefer and prof. S.~Bonazzola for useful
discussions and comments on the basic ideas of the present article,
and Dr.~Nedjalka Petkova for her help in translation of the article
\cite{Brillouin}.

This research was supported in part  by the Fulbright Educational
Exchange Program, Grant Number 01-21-01, by the Scientific Found of
Sofia University, Grant Number 3305/2003, by the Special Scientific
Foundation of JINR, Dubna for 2003, the University of Trieste, INFN,
and by the Abdus Salam International  Centre for Theoretical Physics,
Trieste, Italy, the Sofia University Foundation TCPA, BLTP, JINR,
and the Bulgarian Nuclear Regulatory Agency, Grants  2014-2019.
\end{acknowledgments}

\appendix

\section{Direct derivation of Field Equations from the Einstein Equations}

Using the normal field variables, introduced in Section V, one
can write down the nonzero mixed components $G^\mu_\nu$ of
the Einstein tensor in the basic regular gauge ($\bar\varphi(r)\equiv
0$) as follows: \ben G^t_t\!=\!
{{e^{2\varphi_1-4\varphi_2}}\over{\bar\rho^2}}
\Big(2\bar\rho^2\varphi_1^{\prime\prime}\!-\!2\bar\rho^2\varphi_2^{\prime\prime}
\!-\!(\bar\rho\varphi_1^\prime)^2\!+\!(\bar\rho\varphi_2^\prime)^2\!+\!e^{2\varphi_2}
\Big),\nonumber\\
G^r_r\!=\!{{e^{2\varphi_1-4\varphi_2}}\over{\bar\rho^2}}
\Big(\!(\bar\rho\varphi_1^\prime)^2\!-\!(\bar\rho\varphi_2^\prime)^2\!+
\!e^{2\varphi_2}\Big),
\nonumber\hskip 2.5truecm\\ G^\theta_\theta=
G^\phi_\phi=-{{e^{2\varphi_1-4\varphi_2}}\over{\bar\rho^2}}
\Big(\bar\rho^2\varphi_2^{\prime\prime}\!+\!(\bar\rho\varphi_1^\prime)^2\!-
\!(\bar\rho\varphi_2^\prime)^2\Big).\hskip 1.truecm
\la{Gmunu}\een

Taking into account that the only nonzero component of the
energy-momentum tensor is $T^t_t$ and combining the Einstein
equations (\ref{Einst}), one immediately obtains the field
equations (\ref{FEq0}) and the constraint (\ref{epsilon}). A
slightly more general consideration with an arbitrary rho-gauge
fixing function $\bar\varphi(r)$ yields equations (\ref{FEq})
and constraint(\ref{constraint}).

A similar derivation produces equations (\ref{EL}), which
present another version of the field equations in spherically
symmetrical space-times. In this case: \ben
G^t_t\!=\!{1\over{-g_{rr}}}\left(\!-2\left(\!{{\rho^\prime}\over{\rho}}\!\right)^{\!\prime}
\!-\!3\left(\!{{\rho^\prime}\over{\rho}}\!\right)^{\!2}\!+
\!2{{\rho^\prime}\over{\rho}}{{\sqrt{-g_{rr}}^{\,\prime}}\over{\sqrt{-g_{rr}}}}
\right)\!+\!{{1}\over{\rho^2}},\hskip .3truecm \nonumber\\
G^r_r\!=\!{1\over{-g_{rr}}}\left(\!-\!\left(\!{{\rho^\prime}\over{\rho}}\!\right)^{\!2}\!+
\!2{{\rho^\prime}\over{\rho}}{{\sqrt{g_{tt}}^{\,\prime}}\over{\sqrt{g_{tt}}}}
\right)\!+\!{{1}\over{\rho^2}},\hskip 2.2truecm  \\
G^\theta_\theta\!=\! G^\phi_\phi\!=\!{1\over{-g_{rr}}}
\Bigg(\!\!-\!\left(\!{{\rho^\prime}\over{\rho}}\!\right)^{\!\prime}\!
\!-\!\left(\!{{\rho^\prime}\over{\rho}}\!\right)^{\!2}\!
\!-\!{{\rho^\prime}\over{\rho}}{{\sqrt{g_{tt}}^{\,\prime}}\over{\sqrt{g_{tt}}}}
\!+\hskip 1.4truecm \nonumber
\\
\!{{\rho^\prime}\over{\rho}}{{\sqrt{-g_{rr}}^{\,\prime}}\over{\sqrt{-g_{rr}}}}\!-\!
\left(\!{{\sqrt{g_{tt}}^{\,\prime}}\over{\sqrt{g_{tt}}}}\!\right)^{\!\prime}\!
\!-\!\left(\!{{\sqrt{g_{tt}}^{\,\prime}}\over{\sqrt{g_{tt}}}}\!\right)^{\!2}\!
\!+\!{{\sqrt{g_{tt}}^{\,\prime}}\over{\sqrt{g_{tt}}}}
{{\sqrt{-g_{rr}}^{\,\prime}}\over{\sqrt{-g_{rr}}}}
\Bigg).\nonumber \la{Gmunu_g}\een

\section{Invariant Delta Function for the Weyl Solution}

The function $\rho_{{}_W}(r)\!=\!{{1}\over
4}\!\left(\!\sqrt{r/{\rho_G}}\!+\!\sqrt{{\rho_G}/
r}\right)^2\!\geq\!\rho_G$ that describes the Weyl rho-gauge
fixing is a one-valued function of $r\in (0,\infty)$, but its
inverse function $r_{{}_W}(\rho)$ is a two-valued one:
$r_{{}_W}(\rho)=r_{{}_W}^\pm(\rho)=\rho\left(1\pm\sqrt{1-\rho_G/\rho}\right)^2$,
$\rho\in[\rho_G,\infty)$. Obtaining information about distant
objects only in the optical way, i.e. by measuring only the luminosity
distance $\rho$, one is not able to choose between the two
branches of this function and, therefore, has to consider both
of them as possible results of the observations. Hence, the
3D space ${\cal M}^{(3)}_{{}_W}\{g_{mn}({\bf r })\}$, which
appears in the Weyl gauge, has to be considered as a two-sheeted
Riemann surface with a branch point at the 2D sphere
$\rho^*=\rho_G$ in ${\cal M}^{(3)}_{{}_W}\{g_{\tilde m \tilde
n}({\rho,\tilde \theta,\tilde \phi})\}$ -- the "observable" space
with new spherical coordinates ${\rho,\tilde \theta,\tilde \phi}$.

Due to the limits $r_{{}_W}^+(\rho)\to \infty$ and
$r_{{}_W}^-(\rho)\to 0$ for $\rho\to \infty$, one can write down
the invariant $\delta$-function in the case of the Weyl gauge in the
form \ben
\delta_{g,{}_W}(\rho)={{\sqrt{1-\rho_G/\rho}}\over{4\pi}\rho^4}\,
\delta\!\left({1\over\rho}\right).\la{W_delta_rho}\een

Now taking into account that the equation $1/\rho(r)=0$ has two
solutions: $r=0$ and $1/r=0$ and using the standard formula for
expansion of the distribution  $\delta\big(1/\rho(r)\big)$, we
obtain easily the representation (\ref{Wdelta}).

\section{Independent Nonzero Invariants of Riemann Curvature in Normal Field
Coordinates}
Using the representation (\ref{nc}) of the metric in the normal
field variables and after some algebraic manipulations, one obtains the following
expressions for the possibly independent nonzero invariants of
the Riemann curvature tensor in the problem at hand:

\ben R=-2e^{2(\varphi_1-2\varphi_2)}
\left(\varphi_1^{\prime\prime}-2E_2-E_3 \!\right),\hskip
2.6truecm\nonumber\\ r_1\!=\!{1\over
4}e^{4(\varphi_1\!-\!2\varphi_2)}
\Bigg(\!2\Big(\varphi_1^{\prime\prime}\!-\!E_2\!-\!E_3\Big)^2\!+\!
\Big(\varphi_1^{\prime\prime}\!+\!E_3\Big)^2\Bigg),\hskip
.4truecm\nonumber\\ w_1\!={{C^2}\over {6}},\,\,\,
w_2\!=-{{C^3}\over {36}},\hskip 4.7truecm \la{invars}\\
m_3\!=\!{{C^2}\over{12}}e^{4(\varphi_1\!-\!2\varphi_2)}
\Bigg(\!\Big(\varphi_1^{\prime\prime}\!-\!E_2\!-\!E_3\Big)^2\!\!+\!
2\Big(\varphi_1^{\prime\prime}\!+\!E_3\Big)^2\Bigg),\hskip
.15truecm\nonumber\\
m_5\!=\!{{C^3}\over{36}}e^{4(\varphi_1\!-\!2\varphi_2)}
\Bigg(\!{1\over 2}\!
\Big(\varphi_1^{\prime\prime}\!-\!E_2\!-\!E_3\Big)^2\!\!-\!
2\Big(\varphi_1^{\prime\prime}\!+\!E_3\Big)^2\Bigg),\nonumber\een
where $E_2:=\varphi_2^{\prime\prime} - e^{2\varphi_2}/\bar\rho^2$
and
$E_3:=\varphi_1^\prime{}^2-\varphi_2^\prime{}^2+e^{2\varphi_2}/\bar\rho^2$.

As we can see, the invariants $w_{1,2}$ and $m_5$ in our problem are
real. Hence, in it we may have at most 6 independent invariants.

In addition $(w_1)^3=6\,(w_2)^2$ and the metric (\ref{nc}) falls
into the class II of the Petrov classification \cite{CMcL}. As a
result, the number of independent invariants is at most 5.

Instead of the invariants $w_1$ and/or $w_2$, we use as an independent
invariant
\ben C:=-e^{2\varphi_1\!-\!4\varphi_2}
\Big(2\varphi_1^{\prime\prime}\!+\!
6\varphi_1^{\prime}\left(\varphi_1^{\prime}\!-\!\varphi_2^{\prime}\right)
\!-\!E_2\!-\!2E_3\Big). \la{C}\een
It has the following advantages:

i) It is linear with respect of the derivative
$\varphi_{1}^{\prime\prime}$. This property makes the
expression (\ref{C}) meaningful in the cases when
$\varphi_{1}^{\prime\prime}$ is a distribution;

ii) It is a homogeneous function of first degree with respect to
the Weyl conformal tensor $C_{\mu\nu,\lambda}{}^\xi$;

iii) It is proportional to an eigenvalue of a proper tensor, as
pointed, for example, by Landau and Lifshitz \cite{books}.

iv) The difference between the scalar curvature $R$ and  the
invariant (\ref{C}) yields a new invariant \ben
D\!:=\!R\!-\!C\!=\!3\, e^{2\varphi_1\!-\!4\varphi_2}
\Big(2\varphi_1^{\prime}\big(\varphi_1^{\prime}\!-\!\varphi_2^{\prime}\big)
+E_2 \Big)\,\,\,\,\la{D}\een, which does not include the derivative
$\varphi_1^{\prime\prime}$.

Due to this property, the invariant $D$
in the case of regular solutions  (\ref{sol_f}),
(\ref{sol_NewGauge}) will not contain the Dirac $\delta$-function.
One can use the new invariant $D$ as an independent one, instead
of the invariant $C$.

As seen from Eq. (\ref{invars}), the following functions of
the invariants $r_1$, $m_3$ and
$m_5$:
\ben \, e^{2(\varphi_1\!-\!2\varphi_2)}
\Big(\varphi_1^{\prime\prime} +E_3
\Big)=\sqrt{4\left(m_3/w_1-r_1\right)/3},\hskip .35truecm
\nonumber
\\
e^{2(\varphi_1\!-\!2\varphi_2)}\Big(2\varphi_1^{\prime\prime}\!-\!E_2\!-\!E_3
\Big)\!=\!\sqrt{2\left(4r_1\!-\!m_3/w_1\right)/3}\hskip .5truecm
\la{m5}\een are linear with respect to the derivative
$\varphi_1^{\prime\prime}$. Using proper linear combinations of
these invariants and the scalar curvature $R$, one can easily check
that the quantities $E_2\,e^{2(\varphi_1\!-\!2\varphi_2)}$ and
$E_3\,e^{2(\varphi_1\!-\!2\varphi_2)}$ are independent invariants,
too.

Thus, our final result is that, in general, for the metric
(\ref{nc}) in the basic regular gauge $\bar\varphi\equiv 0$ the
Riemann curvature tensor has the following four independent
invariants: \ben
I_1:=e^{2(\varphi_1\!-\!2\varphi_2)}\varphi_1^{\prime\prime},\hskip
3.77truecm\nonumber\\
I_2:=e^{2(\varphi_1\!-\!2\varphi_2)}\left(\varphi_2^{\prime\prime}-
e^{2\varphi_2}/\bar\rho^2\right),\hskip 1.73truecm\\
I_3:=e^{2(\varphi_1\!-\!2\varphi_2)} \left(-\varphi_1^\prime{}^2+
\varphi_2^\prime{}^2-e^{2\varphi_2}/\bar\rho^2
\right)/2,\nonumber\\
I_4:=e^{2(\varphi_1\!-\!2\varphi_2)}\varphi_1^{\prime}
\big(\varphi_1^{\prime}\!-\!\varphi_2^{\prime}\big)/2. \hskip
2.03truecm\nonumber \la{I}\een These invariants are linear with
respect to the second derivatives of the functions $\varphi_{1,2}$
-- a property, which is of critical importance when we have to
work with distributions $\varphi_{1,2}^{\prime\prime}$.


\begin{thebibliography}{}
%%%%%%%%%%%%%%%%%%%%%%%%%%%%%%%%%%%%%%%%%%%%%%%%%%%%%%%%%%%%%%%%%%%%%%%%%%%%%%%
\bibitem{LOGOOPEN} https://losc.ligo.org

\bibitem{Barack2018}Leor Barack et al, {\em Black holes, gravitational waves and fundamental
physics: a roadmap}, arXiv:1806.05195.

%
\bibitem{Schwarzschild} K.~Schwarzschild, Sitzungsber. Preus.
Acad. Wiss. Phys. Math. Kl., 189 (1916).
%QNM    Misner C W 1963 Ann. Phys. 24 102–17
%
\bibitem{ML} J. Mitchell, Phil. Trans. R. Soc. London, {\bf 74}, 35-37 (1784).
             P.S. Laplace, Laplace, {\em Exposition du Système du Monde} Part 11. Paris (1796).
%
\bibitem{Starobinskij1973} A. A. Starobinskij and S. M. Churilov, Zhurnal Eksperimentalnoi i Teoreticheskoi Fiziki {\bf 65} 3 (1973).
%
\bibitem{ChandraDetweiler1975} S. Chandrasekhar and S. L. Detweiler, Proc. Roy. Soc. Lond. A {\bf 344}  441-452 (1975).

\bibitem{Chandrasekhar1983} S. Chandrasekhar, The Mathematical Theory of Black Holes, Oxford University Press, New York (1983).

\bibitem{Nagar05}A. Nagar, L, Rezzolla,  Class. Quantum Grav. {\bf 22}  R167 (2005)


\bibitem{Kokkotas1999} K.D. Kokkotas, B.G. Schmidt, LivingRew.Rel.2:2(1999).

\bibitem{Berti06} E. Berti, V. Cardoso, C. M. Will, PRD {\bf 73} 064030 (2006).


\bibitem{Berti09} E. Berti, V. Cardoso, and A. O. Starinets, Class. Quantum Grav. {\bf 26}  163001 (2009).

\bibitem{Nigel2016} Nigel T. Bishop · Luciano Rezzolla, (\em Extraction of gravitational waves in numerical relativity) Living Rev Relativ 19:2 (2016).

\bibitem{Testing2019} Maximiliano Isi, Matthew Giesler, Will M. Farr, Mark A. Scheel, Saul A. Teukolsky
     {\em  Testing the no-hair theorem with GW150914}, arXiv:1905.00869.

%Echos

\bibitem{Cardoso16} V. Cardoso, E. Franzin, P. Pani, Phys. Rev. Lett. 116, 171101 (2016).

\bibitem{Mark2017} Zachary Mark, Aaron Zimmerman, Song Ming Du, Yanbei Chen,{\em A recipe for echoes from exotic compact objects}, arXiv:1706.06155.

\bibitem{Cardoso2017} V. Cardoso, P. Pani,  {\em The observational evidence for horizons: from echoes to precision gravitational-wave physics}, arXiv:1707.03021.

\bibitem{Conklin18}Randy S. Conklin, Bob Holdom, and Jing Ren, {\em Gravitational wave echoes through new windows}, arXiv:1712.06517.

\bibitem{Wang2018} Qingwen Wang, Niayesh Afshordiy, {\em Black Hole Echology: The Observer's Manual}, arXiv:1803.02845.

\bibitem{Raposo18} G. Raposo, P. Pani1, M. Bezares, C. Palenzuela, V. Cardoso,
             {\em Anisotropic stars as ultracompact objects in General Relativity}, arXiv:1811.07917 (2018).

\bibitem{Cardoso2019} V. Cardoso, P. Pani, {Testing the nature of dark compact objects: a status report}, Living Reviews in Relativity (2019). https://doi.org/10.1007/s41114-019-0020-4

\bibitem{EHT} The Event Horizon Telescope Collaboration,

The Astrophysical Journal Letters, 875:L1 (2019);

The Astrophysical Journal Letters, 875:L2 (2019);

The Astrophysical Journal Letters, 875:L3 (2019);

The Astrophysical Journal Letters, 875:L4 (2019);

The Astrophysical Journal Letters, 875:L5 (2019);

The Astrophysical Journal Letters, 875:L6 (2019).

\bibitem{Fiziev2003} P. P. Fiziev {\em Gravitational Field of Massive Point Particle in General Relativity} arXiv:gr-qc/0306088 v12; Abdus Salam ICTP, Trieste, Italy, preprint IC/2003/122;

 \bibitem{Fiziev2004a} P. P. Fiziev {\em On the Solutions of Einstein Equations with Massive Point Source}, gr-qc/0407088;
                     In Proceedings of Conference GAS04, Kiten 2004,Editors: P. Fiziev, M. Todorov, St. Kliment Ohridski University Press, 2005, pp. 70-83.

                     Fiziev P. P., The Gravitational Field of Massive Non-Charged Point Source
                     in General Relativity, gr-qc/0412131; In Proceedings of Conference Modern Trends
                     in Theoretical Physics, Gravity and Astrophysics, dedicated to prof. Georgi Manev,
                     Sofia 2004, pp. 239-250.

%
\bibitem{Petrov2018} Alexander N. Petrov, General Relativity and  Gravitation {\bf 50} 6 (2018).
%
\bibitem{books}   L.~D.~Landau, E.~M.~Lifshitz, {\em The Classical Theory of
                  Fields}, 2d ed.; Reading, Mass: Addison-Wesley, 1962;
                  V.~A.~Fock, {\em The Theory of Space, Time and
Gravitation},
                  Pergamon, Oxford, 1964.
                  R.~C.~Tolman, {\em Reativity, Thermodynamics and
                  Cosmology}, Claderon Press, Oxford, 1969.
                  S.~Weinberg,{\em Gravitation and Cosmology}, Wiley,
                  N.Y., 1972;
                  C.~Misner, K.~S.~Thorne, J.~A.~Wheeler, {\em Gravity},
                  W.~H.~Freemand\&Co.,1973.
%
\bibitem{Exact}     D.~Kramer, H.~Stephani, M.~Maccallum,
                    E.~Herlt, Ed.~E.~Schmutzer, {\em Exact Solutions of the Einstein
                    Equations}, Deutscher Verlag der
                    Wissenschaften, Berlin, 1980.
                    H.~Stephani, D.~Kramer, M.~Maccallum,
                    E.~Herlt, {\em Exact Solutions of Einstein's
                    Field Equations}, Sec. Ed., Cambridge University
                    Press, 2003.
%
\bibitem{GS} M.~G\"ockeler, T.~Sch\"ucker, {\em Differential Geometry, Gauge
Theories, and Gravity}, Cambridge University Press, Cambridge, 1987.
%
\bibitem{Rubakov} V. Rubakov, {\em Classical Theory of Gauge Fields},
Princeton University
                  Press, Princeton and Oxford, 1999.
%
\bibitem{Eddington}  A.~S.~Eddington, {\em The mathematical theory of
relativity}, 2nd ed. Cambridge, University Press, 1930 (repr.1963).
%
%
\bibitem{Hilbert} D.~Hilbert, Nachr. Ges. Wiss. G\"otingen, Math.
                  Phys. Kl., 53 (1917).
%
\bibitem{GH}  R.~Gautreau, B.~Hoffmann, Phys. Rev. D{\bf 17}, 2552 (1978).
%
\bibitem{Gelfand}   L.~Schwartz, {\em Th\'eorie des distributions}
                    I, II, Paris, 1950-51;\,
                    I.~M.~Gel'fand, G.~E.~Shilov, {\em Generalized
                    Finctions}, N.Y., Academic Press, 1964;\,
                    H.~Bremermann, {\em Distrinutions, Complex Variables and Fourier
                    Transform}, Addison-Wesley Publ. Co. Reading,
                    Massachusetts, 1965.
%
\bibitem{ComJann} J.~T.~Combridge, Phil. Mag. {\bf 45}, 726
                  (1923);
                  H, Janne, Bull. Acad. R. Belg. {\bf 9}, 484 (1923).
                  M.Kpher, Zeitschrift fiir Physik, Bd. 134, S. 286--305 (1953); Zeitschrift f/it Physik, Bd. 134, S. 306--316 (t953).
                  F.J.Belinfante, Phys.~Rev. {\bf 98}, 793 (1955).
                  Ya. I. Pugachev and V. D. Gun'ko, Soviet Phys. Journ.,
                  Izvestiya Vysshikh Uchebnykh Zavednii, Fizika, {\bf 10}, 46, (1974).
                  D.~H.~Menzel, M\'em.~Soc.~Roy.~Sci.~Li\'ege, {\bf9}, 343 (1976).

%
\bibitem{AL} S.~Antoci, D.-E.~Liebscher, Astron. Nachr. {\bf 322}, 137
(2001);
                       S.~Antoci, D.-E.~Liebscher, L.~Mihich,
                       Class. Quant. Grav. {\bf 18}, 3466 (2001).
%
\bibitem{FN} V.~P.~Frolov, I.~D.~Novikov, {\em Black Hole
             Physics}, Kluwer Acad. Publ., 1998.
%
\bibitem{Droste} J.Droste, Proc.K.Ned.Acad.Wet., Ser., A{\bf 19}, 197
(1917).
%
\bibitem{Weyl} H.~Weyl, Ann. Phys., Leipzig, {\bf 54}, 117 (1917).
%
\bibitem{Einstein} A.~Einstein, N.~Rosen, Phys. Rev., {\bf 48}, 73 (1935).
%
\bibitem{Bel1} Ll.~Bel, J.~Math.~Phys. {\bf 10}, 1501 (1969);
                        Gen. Rel. and Grav. {\bf 1}, 337 (1971);
               Ll.~Bel., J.~Llosa, Gen. Rel. and Grav. {\bf 27}, 1089
(1995);
               Ll.~Bel,  Gen. Rel. and Grav. {\bf 28}, 1139 (1996);
%
\bibitem{Bel2} J.~M.~Aguirregabiria, Ll.~Bel, J.~Martin, A.~Molina, E.~Ruiz
               gr-qc/0104019, Gen. Rel. and Grav. {\bf 33} 1809 (2001);
               J.~M.~Aguirregabiria, Ll.~Bel, gr-qc/0105043,
               Gen. Rel. and Grav. {\bf 33}, 2049 (2001) ;
               Ll.~Bel, gr-qc/0210057.
%
\bibitem{Fiziev2004} P. P. Fiziev, {\em Novel Geometrical Models of Relativistic Stars. I. The General Scheme}, arXiv:astro-ph/0409456;
                      P. P. Fiziev, {\em Novel Geometrical Models of Relativistic Stars. II. Incompressible Stars and Heavy Black Dwarfs}, arXiv:astro-ph/0409458;
                      P. P. Fiziev, {\em Novel Geometrical Models of Relativistic Stars III. The Point Particle Idealization}, arXiv:astro-ph/0409610.
%
\bibitem{ADM} E.~Arnowitt,x
 S.~Deser, C.~M.~Misner, Phys. Rev. Lett.
              D{\bf 4}, 375 (1960); Phys. Rev. {\bf 120} 321 (1960);
              in {\em Gravitation: an introduction to current research},
              Ed. L.~Witten, John Wiley, N.Y., 1962.
%
\bibitem{Lichnerowicz} A.~Lichnerowicz, {\em Th\'eories Relativistes de la
Gravitation
         et de l'Electromagn\'etisme}, Paris, Masson, 1955.
%
\bibitem{YB} Y.~Bruhat, {Cauchy Problem}, in {\em Gravitation: an
introduction to current research}, ed. L. Witten, John Wiley, N.Y. , 1962.
%
\bibitem{Abrams}  L.~S.~Abrams, Phys.~Rev.~D{\bf 20}, 2474 (1979);
                  Can.~J.~Phys., {\bf 67}, 919 (1989);
                  Physica {\bf A}, 131 (1996);
                  Int.~J.~Theor.~Phys. {\bf 35}, 2661 (1996).
%
\bibitem{Wheeler1955} A.~Einstein, N.~Rosen, Phys. Rev., {\bf 48}, 73 (1935).

                  J.~A.~Wheeler, Phys. Rev. {\bf 97}, 511 (1955).

                  C.~W.~Misner, Phys. Rev. {\bf 118}, 1110 (1960).

                  J.~A.~Wheeler, Rev. Mod. Phys.  {\bf 33}, 63 (1961).

                  C.~W.~Misner, Ann. Phys. {\bf 24}, 102 (1963).

                  R.~W.~Lindquist, J. Math. Phys. {\bf 4}, 938 (1963).

                  D.~R.~Brill, R.~W.~Lindquist, Phys. Rev. {\bf
                  131} 471 (1963).

                  J.~Bowen, J.~W.~York, Phys. Rev. D{\bf 21}, 2047 (1980).

                  V.~P.~Frolov, I.~D.~Novikov, {\em Black Hole
                  Physics}, Kluwer Acad. Publ., 1998
%
\bibitem{RW} Regge T and Wheeler J A, Phys. Rev. 108 1063 (1957).
%
\bibitem{Z} Zerilli F J, Phys. Rev. D 9 860 (1974).
%
\bibitem{Chandrasekhar1983} S. Chandrasekhar, The Mathematical Theory of Black Holes (Oxford University Press, Oxford, 1983).
%
\bibitem{Fiziev2006} Plamen P Fiziev, CQG. {\bf 23}, 2447–2468 (2006).
%
\bibitem{Fiziev2009} Plamen P. Fiziev, PRD {\bf 80}, 124001 (2009).
%
\bibitem{Fiziev2011} Plamen Fiziev, Denitsa Staicova, PRD {\bf 84}, 127502 (2011).
%
\bibitem{Cartan} Cartan E 1983 {\em Geometry of Riemannian Spaces}, Math. Sci. Press, Brookline, Massachusetts

                 Brans C H, Jour. Math. Phys {\bf 6} 94-102, (1965).

                 Karlhede A, Lindstr\"om U, Aman J E, Gen. Rel. and Grav. {\bf 14} 569 (1982).

                 Skea JEF,  Clas. Quant. Grav. {\bf 14} 2947-2950 (1997).

                 Skea JEF,  Clas. Quant. Grav. {\bf 17} L69-L74, (2000).

                 Lake K, Gen. Rel. and Grav. {\bf 36} 1159, (2004).

                 Coley A, Hervik S, Pelavas N,  Class. Quant. Grav. {\bf 26} 025013, (2009).
%
\bibitem{Karlhede1982} Karlhede A, Lindstrom U,  Etan J, General Relativity and Gravitation, Vol. 14, No. 6 (1982).
%
\bibitem{Fiziev2010} P P Fiziev, Class. Quantum Grav. {\bf 27}, 135001 (2010).
%
\bibitem{CMcL} J.~Carminati, R.~G.~McLenaghan, J.~Math.~Phys. {\bf32}, 3135 (1991).
%
\bibitem{GT} R.~Geroch, J.~Traschen, Phys. Rev. D{\bf 36} 1017 (1987).
%
\bibitem{TaubRaju} A.~H.~Taub, J. Math. Phys. {\bf 21} 1423 (1979).
                   C.~K.~Raju, J. Phys. A: Math. Gen. {\bf 15} 1785 (1982).
%
\bibitem{Dirac} P.~A.~M.Dirac, Proc. Roy Soc. (London), A{\bf 270}, 354 (1962);
                              Conference in Warszawa and Jablonna, L. Infeld ed., Gauthier-Villars, Paris (1964), pp. 163-175.
%
\bibitem{Brillouin} M.~Brillouin, Le Journal de physique et Le Radium {\bf 23}, 43 (1923).
%
\bibitem{MM} P.~O.~Mazur, E.~Motolla, {\em Gravitational Condensate Stars:
             An Alternative to Black Holes}, gr-qc/0109035;
             M.~Visser, D.~L.~Wiltshire, {\em Stable Gravistars - an Alternative
             to Black Holes}, gr-qc/0310107.
%
\bibitem{Poincare} H.~Poincare, Science et M\'etode, Book 1, Flammarion, Paris, 1934.
%
\bibitem{Fiziev2007} P.P. Fiziev, {\em Static Fundamental Solutions of Einstein Equations
            and Superposition Principle in Relativistic Gravity} arXiv:gr-qc/0701108.

\bibitem{FizievDimitrov2004} P. P. Fiziev, S. V. Dimitrov, {\em  Point Electric Charge in General Relativity}            arXiv:hep-th/0406077
%
\bibitem{AKL} M.~A.~Abramowicz, W.~Klu\'zniak, J-P.~Lasota, Astron. Astrophys., {\bf 396}, L31 (2002); astro-ph/0207270.
%
\bibitem{RN} E.~W.~Mielke, F.~E.~Schunck, Nucl. Phys. B{\bf 564}, p.185 (2000);
             gr-qc/0001061.
R.~Narayan, {\em Evedences for the Black Hole Event Horison}, astro-ph/0310692.
%
\bibitem{LATOR} S.~G.~Turyshev, J.~G.~Williams, K.~Nordtvedt, Jr., M.~Shao,
                T.~W.~Murphy, Jr., {\em 35 Years of Testing Relativistic
                Gravity: Where do we go from here?}, gr-qc/0311039.
%
\bibitem{Jafari2019} A. Jafari, {\em GRAVITATIONAL RADIATION FROM BINARIES: A PEDAGOGICAL INTRODUCTION}, arXiv:1908.04410.
%
\bibitem{Buchdahl} Buchdahl H.A, Phys Rev 116:1027 (1959); PhysRev.116.1027 (1959).

\end{thebibliography}
\end{document}